\documentclass[10pt,journal,compsoc]{IEEEtran}
\usepackage{multirow}
\usepackage[usenames,dvipsnames]{color}
\usepackage[table]{xcolor}
\usepackage{amssymb}% http://ctan.org/pkg/amssymb
\usepackage{pifont}% http://ctan.org/pkg/pifont

\ifCLASSOPTIONcompsoc
  \usepackage[nocompress]{cite}
\else
  \usepackage{cite}
\fi
\ifCLASSINFOpdf
  \usepackage[pdftex]{graphicx}
\else
\fi
\usepackage{amsmath}
\usepackage{mathptmx}
\usepackage{mathtools}

\usepackage{float}

\newcommand{\tr}{\text{tr}}
\newcommand{\diam}{\text{diam}}
\newcommand{\den}{\text{den}}
\newcommand{\triangles}{\text{triangles}}
\newcommand{\mean}{\text{mean}}
\newcommand{\dist}{\text{dist}}
\newcommand{\connectedtriples}{\text{connected triples}}

\newcommand{\hang}{\textcolor{black}}
\newcommand{\vahan}{\textcolor{black}}

\newcommand{\ross}{\textcolor{black}}

\usepackage{xpatch}
\makeatletter
\patchcmd\@makecaption{\\}{.~}{}{\fail}
\makeatletter
\usepackage{array}
\usepackage{wrapfig}
\usepackage{paralist}
\usepackage{cancel}
\usepackage{hyperref}
\begin{document}
%
% paper title
% Titles are generally capitalized except for words such as a, an, and, as,
% at, but, by, for, in, nor, of, on, or, the, to and up, which are usually
% not capitalized unless they are the first or last word of the title.
% Linebreaks \\ can be used within to get better formatting as desired.
% Do not put math or special symbols in the title.
\title{Same Stats, Different Graphs:  Exploring the Space of Graphs in Terms of Graph Properties\thanks{This is a journal version of a paper that appeared in the proceedings of the 26th Symposium on Graph Drawing and Network Visualization (GD'18).}}%, which was selected by the GD program chairs and invited for publication in TVCG.}}
%
%
% author names and IEEE memberships
% note positions of commas and nonbreaking spaces ( ~ ) LaTeX will not break
% a structure at a ~ so this keeps an author's name from being broken across
% two lines.
% use \thanks{} to gain access to the first footnote area
% a separate \thanks must be used for each paragraph as LaTeX2e's \thanks
% was not built to handle multiple paragraphs
%
%
%\IEEEcompsocitemizethanks is a special \thanks that produces the bulleted
% lists the Computer Society journals use for "first footnote" author
% affiliations. Use \IEEEcompsocthanksitem which works much like \item
% for each affiliation group. When not in compsoc mode,
% \IEEEcompsocitemizethanks becomes like \thanks and
% \IEEEcompsocthanksitem becomes a line break with idention. This
% facilitates dual compilation, although admittedly the differences in the
% desired content of \author between the different types of papers makes a
% one-size-fits-all approach a daunting prospect. For instance, compsoc 
% journal papers have the author affiliations above the "Manuscript
% received ..."  text while in non-compsoc journals this is reversed. Sigh.
\author{Hang Chen\thanks{The University of Arizona, Tucson, AZ, USA} \and Utkarsh Soni\thanks{Arizona State University, Tempe, AZ, USA} \and Yafeng Lu\footnotemark[3] \and Vahan Huroyan\footnotemark[2] \and Ross Maciejewski\footnotemark[3] \and Stephen Kobourov\footnotemark[2]}

\author{Hang~Chen,
        Utkarsh~Soni,
        Yafeng~Lu,
        Vahan~Huroyan,
        Ross~Maciejewski,
        Stephen~Kobourov% <-this % stops a space
\IEEEcompsocitemizethanks{\IEEEcompsocthanksitem H. Chen, V. Huroyan, S. Kobourov are with the Department
of Computer Science at The University of Arizona, Tucson, AZ, USA.\protect
\IEEEcompsocthanksitem U. Soni, Y. Lu, R. Maciejewski are with School of Computing, Informatics, and Decision Systems Engineering at Arizona State University, AZ, USA.}% <-this % stops an unwanted space
%\thanks{Manuscript received April 19, 2005; revised August 26, 2015.}
}

\IEEEtitleabstractindextext{%
\begin{abstract}
Data analysts commonly utilize statistics to summarize large datasets. While it is often sufficient to explore only the summary statistics of a dataset (e.g., min/mean/max), Anscombe's Quartet demonstrates how such statistics can be misleading. We consider a similar problem in the context of graph mining.
%similar problem in that graph properties (e.g., density, connectivity, clustering coefficient) may not capture all of the critical properties of a given graph. 
To study the relationships between different graph properties, we examine 
low-order non-isomorphic graphs and provide a simple visual analytics system to explore correlations across multiple graph properties. However, for larger graphs,  studying the entire space quickly becomes intractable. We use different random graph generation methods to further look into the distribution of graph properties for higher order graphs and investigate the impact of various sampling methodologies.
We also describe a method for generating many graphs that are identical over a number of graph properties and statistics yet are clearly different and identifiably distinct.
\end{abstract}

% Note that keywords are not normally used for peerreview papers.
}

% make the title area
\maketitle

% To allow for easy dual compilation without having to reenter the
% abstract/keywords data, the \IEEEtitleabstractindextext text will
% not be used in maketitle, but will appear (i.e., to be "transported")
% here as \IEEEdisplaynontitleabstractindextext when the compsoc 
% or transmag modes are not selected <OR> if conference mode is selected 
% - because all conference papers position the abstract like regular
% papers do.
\IEEEdisplaynontitleabstractindextext
% \IEEEdisplaynontitleabstractindextext has no effect when using
% compsoc or transmag under a non-conference mode.

% For peer review papers, you can put extra information on the cover
% page as needed:
% \ifCLASSOPTIONpeerreview
% \begin{center} \bfseries EDICS Category: 3-BBND \end{center}
% \fi
%
% For peerreview papers, this IEEEtran command inserts a page break and
% creates the second title. It will be ignored for other modes.
\IEEEpeerreviewmaketitle

\begin{figure}[H]
    \centering
	\includegraphics[width=.43\textwidth]{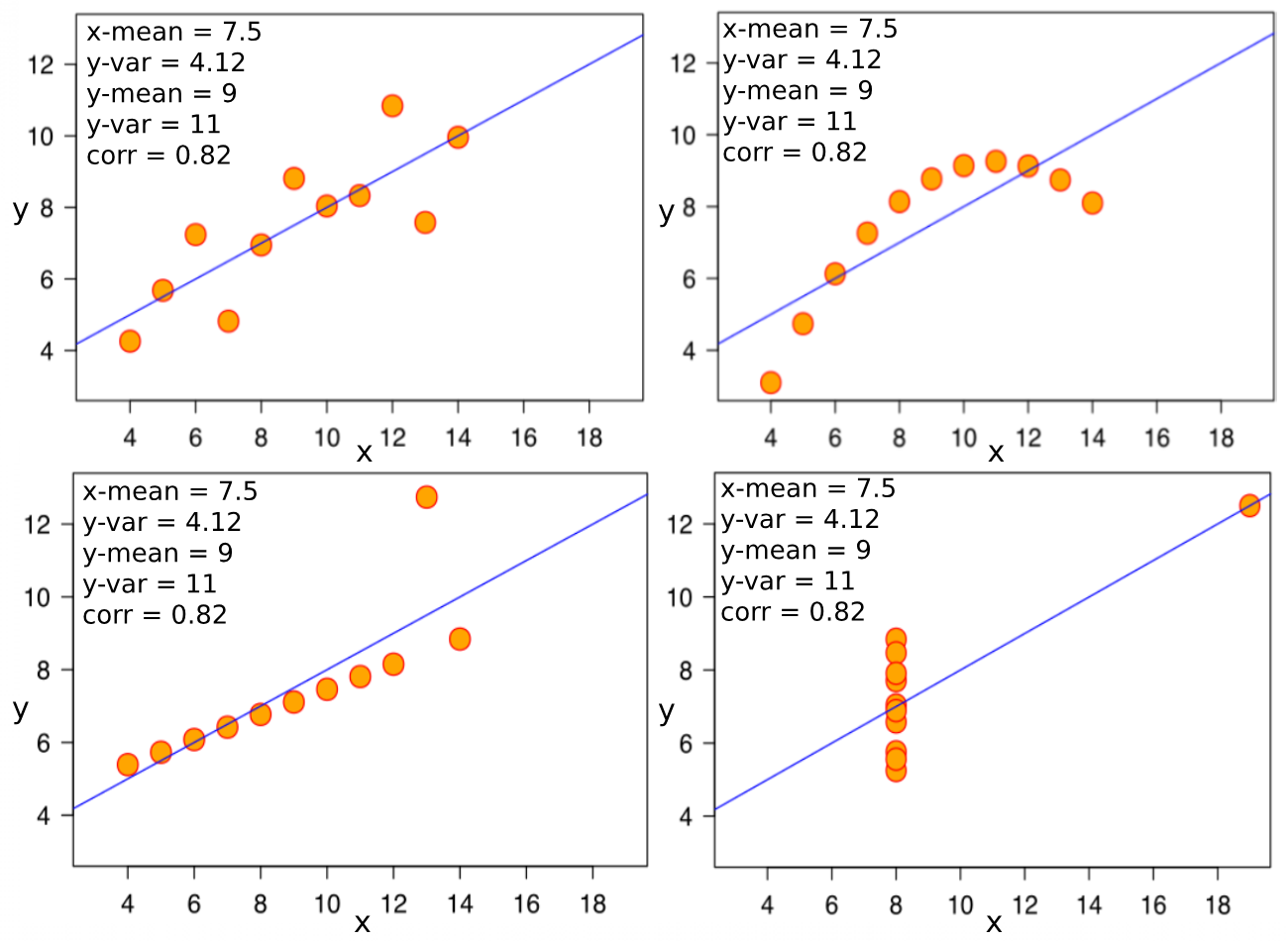}
	\vspace{-0.4cm}
	\caption{Anscombe's quartet: All four datasets have the same mean and standard deviation in $x$ and $y$ and $(x,y)$-correlation.\label{fig:AC}}
% 	\caption{Anscombe's quartet.\label{fig:AC}}
	\centering
\end{figure}

\IEEEraisesectionheading{\section{Introduction}\label{sec:introduction}}
% Computer Society journal (but not conference!) papers do something unusual
% with the very first section heading (almost always called "Introduction").
% They place it ABOVE the main text! IEEEtran.cls does not automatically do
% this for you, but you can achieve this effect with the provided
% \IEEEraisesectionheading{} command. Note the need to keep any \label that
% is to refer to the section immediately after \section in the above as
% \IEEEraisesectionheading puts \section within a raised box.
\IEEEPARstart{S}{tatistics} are often used to summarize a large dataset. In a way, one hopes to find the ``most important" statistics that capture one's data. For example, when comparing two countries, we often specify the population size, GDP, employment rate, etc. The idea is that if two countries have a ``similar" statistical profile, they are similar (e.g., France and Germany have a more similar demographic profile than France and USA). However, Anscombe's quartet~\cite{Anscombe}  convincingly illustrates that datasets with the same values over a limited number of statistical properties can be fundamentally different -- a great argument for the need to visualize the underlying data; see Fig.~\ref{fig:AC}.

\begin{figure}[H]
	\centering
	\fbox{\includegraphics[width=0.19\textwidth]{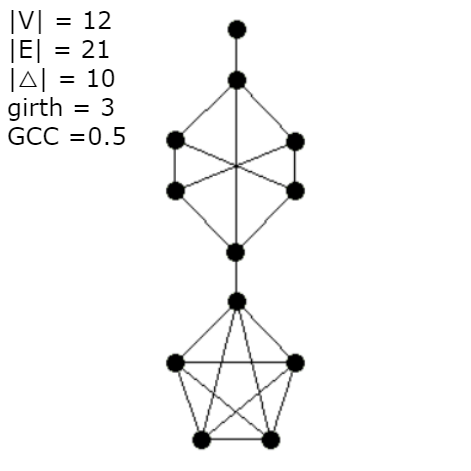}}\fbox{\includegraphics[width=0.19\textwidth]{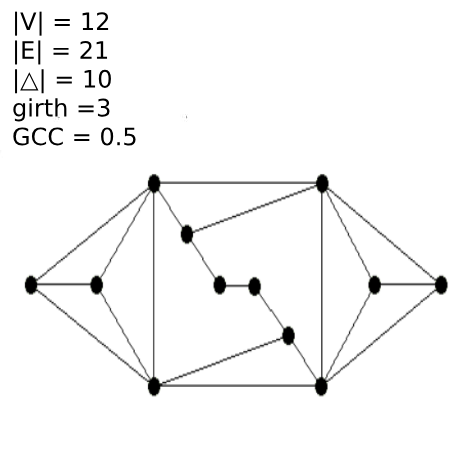}}
    	\fbox{\includegraphics[width=0.19\textwidth]{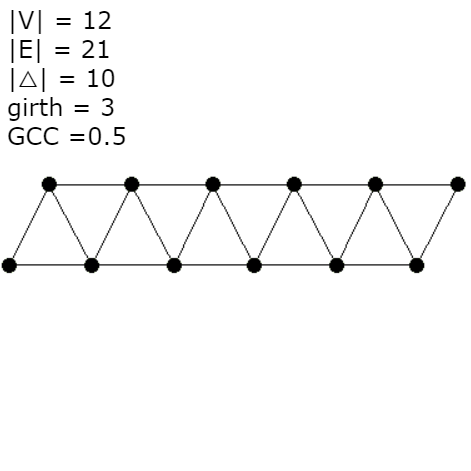}}\fbox{\includegraphics[width=0.19\textwidth]{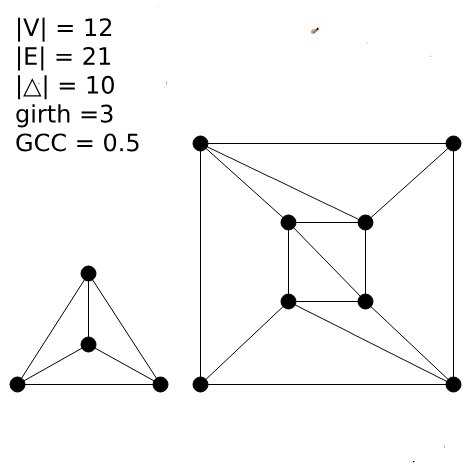}}
	\caption{These four graphs share the same 5 common properties: $|V| = 12$, $|E| = 21$, number of triangles $|\bigtriangleup|=10$, girth $=3$ and global clustering coefficient {\em GCC}$=0.5$.
		However, structurally the graphs are very different: some are planar, others are not, some show regular patterns and are symmetric, others are not, and finally, one of the graphs is disconnected, another is 1-connected and the rest are 2-connected.\label{fig:HC}}
\end{figure}

Similarly, in the graph analytics community, a variety of properties and statistics are being used to summarize graphs, such as graph density, average path length, global clustering coefficient, etc. However, summarizing a graph with a fixed set of graph properties leads to the problem illustrated by Anscombe. 
It is easy to construct several graphs that have the same properties (e.g., number of vertices, number of edges, number of triangles, girth, clustering coefficient) while the underlying graphs are clearly different and identifiably distinct; see Fig.~\ref{fig:HC}. 
From a graph theoretical point of view, these graphs are very different: they differ in  connectivity, planarity, symmetry, and other properties.

Recently, Matejka and Fitzmaurice~\cite{matejka2017same} proposed a dataset generation method that can modify a given 2-dimensional point set (like the ones in Anscombe's quartet) while preserving its summary statistics but  changing its visualization (what they call ``graph"). Given the graphs in Fig.~\ref{fig:HC}, we consider whether it is also possible to modify a given graph and preserve a given set of {\color{black}graph properties} while  changing other graph properties.
Note that the problem is much easier for 2D point sets and statistics, such as mean, deviation and correlation, than for graphs where many graph properties are structurally correlated (e.g., diameter and average path length).
With this in mind, we first consider how can we fix a few graph properties (such as the number of nodes, number of edges, number of triangles) and vary another property (such as the clustering coefficient or connectivity). We find that there is a spectrum of possibilities. Sometimes the ``unrestricted" {\color{black}properties} can vary dramatically, sometimes not, and the outcome depends on two issues: (1) the inherent correlation between some properties (e.g., density and number of triangles), and (2) the bias in graph generators.
%\yafeng{For example, small-word graph generators are in favor of generating large clustering coefficients.}

We begin by studying the correlation between graph {\color{black} properties} across the set of all non-isomorphic graphs with up to 10 vertices. Recall that two vertex-labeled graphs are isomorphic if just by relabeling the vertices of one of the graphs we can obtain the other. Thus, two non-isomorphic graphs must be structurally different. 
% $|V| = 5$ to $|V|=10$ ($|v|$: number of vertices). 
The statistical properties derived for all graphs for a fixed number of vertices provide further information about certain ``restrictions." In other words, the range of one {\color{black} property} may be restricted if another property is fixed. However, we cannot explore the entire space of graph properties and correlations. As the number of vertices grows, the number of different non-isomorphic graphs grows super-exponentially. For $|V|=1, 2 \dots 9$ the numbers are $1, 2, 4, 11, 34, 156, 1044, 12346,$ $274668$, but already for $|V|=16$ we have $6\times 10^{22}$ 
non-isomorphic graphs. 

To go beyond ten vertices, we use graph generators based on 
models, such as Erd\"os-R\'enyi and Watts-Strogatz. However, different graph generators have different biases and these can impact the results. 
We study the extent to which sampling using random generators can represent the whole graph set for an arbitrary number of vertices with respect to their coverage of the graph properties. One way to evaluate the performance of random generators is to compare the generated graphs to the total set of non-isomorphic graphs, which is available for $|V|\leq 10$ vertices. If we randomly generate a small set of graphs (also for  $|V|\leq 10$ vertices) using a given graph generator, we can explore how well the sample and generator cover the space of graph properties. In this way, we can begin exploring the problem ``same stats, different graphs'' for larger graphs. 

We have put together a visual analytics exploration tool for the space of all low-order ($\leq 10$)
non-isomorphic graphs and sampled higher order
graphs. We include a generator for ``same stats, different graphs," i.e., multiple graphs that are identical over a number of graph properties, yet are clearly different. Data and tools are available at \url{http://differentgraphs.cs.arizona.edu}.
This work illustrates the challenges associated with analyzing graphs based solely on sampling summary properties. Overcoming such challenges is critical for network analysis, graph mining, etc., where the analysis of graphs is often done at the graph property level, which can miss critical topological information.

\subsection{Definitions and Conventions}

%\hang{In this paper, We use statistics refer to , measure refer to }

The object of study in this paper are simple, undirected, unweighted graphs, $G=(V,E)$ with $|V|$ nodes and $|E|$ edges. Specifically, we study the set of non-isomorphic graphs, considering each of them as a high-dimensional point. Recall that two graphs $G_1=(V_1, E_1)$ and $G_2=(V_2, E_2)$ are isomorphic if they are identical up to relabeling their vertices; that is, if there exists a bijection $f: V_1 \to V_2$ such that for any $u_1, v_1 \in V_1$ and $\{u_1, v_1\} \in E_1$ if and only if $\{ f(u_1), f(v_1)\} \in E_2$.

Throughout the paper we use the term {\em graph properties} to refer to the properties of a graph (e.g., diameter, density). We also use the term {\em statistics} when describing a graph via a collection of {\color{black}graph properties} (e.g., the values for diameter, density). The 10 properties under consideration are discussed in Section~\ref{sec:graph_stats}. By {\em ground truth} dataset we refer to the dataset in $\mathbb{R}^{10}$ that represents the  set of $10$ properties of all non-isomorphic graphs with fixed number of vertices $|V|$.

Note that the set of all non-isomorphic graphs is not in a bijection with the ground truth dataset in 10 dimensional space. A pair of non-isomorphic graphs does indeed correspond to a pair of distinct points in $\mathbb{R}^{10}$. On the other hand, a pair of non-isomorphic graphs that share the same 10 {\color{black}graph properties} get mapped to the same point in $\mathbb{R}^{10}$, which is an instance of the titular ``same stats, different graphs."

%\vahan{We would like to note that the set of all non-isomorphic graphs is not is not in a bijection with the corresponding 10 dimensional set of graph properties. In fact, any instance of 2 “same stats, different graphs” corresponds to 2 non-isomorphic graphs that get mapped to the same 10D point.}

\subsection{Structure of This Paper}

This paper is organized as follows: Section~\ref{sec:related_work} summarizes related work; Section~\ref{sec:preliminary_findings} presents preliminary experiments and findings about low-order graphs; Section~\ref{sec:finding_same_stat} discusses methods of generating different graphs with same or similar properties; Section \ref{sec:rep_cov} discusses different measures for  estimating graph generators from the point of view of coverage and representation of the underlying space of non-isomorphic graphs;  Section~\ref{sec:rep_cov_exper} compares the coverage and representation of the graph generators used in the paper. We conclude with a brief discussion and directions for future work in section~\ref{sec:conclusion}

\section{Related Work}
\label{sec:related_work}
There is a great deal of related work, but here we focus on graph properties that are studied in the graph analysis and graph mining literature, a review of the major random graph generators, and work on exploration and visualization of graph properties.
%We also consider different graph generators \hang{which generate simple graph}. 

\begin{table*}
\caption{The set of graph \vahan{properties} considered in this paper.} % title of Table
\vspace{-0.2cm}
\label{Table:properties}
%\centering % used for centering table
%\rowcolors{2}{gray!25}{white}
{\renewcommand{\arraystretch}{1.5}%
\begin{tabular}{ m{4cm} m{9cm} m{2.5cm} } % centered columns (4 columns)
    %\rowcolor{gray!50}

\hline\hline %inserts double horizontal lines
%\hline % inserts single horizontal line
  Name & Formula & Reference\\ % inserts table
%heading
\hline % inserts single horizontal line
%clustering coefficient& $c(v) = \frac{|\{(u,w)|u,w \in \Gamma(v) , (u,w) \in E\}|}{|\Gamma(v)|(|\Gamma(v)|-  1)/2}, v,u,w \in V$& ~\cite{chakrabarti2007graph,hanhijarvi2009randomization,mcglohon2011statistical,li2011graph,kairam2012graphprism,chakrabarti2006graph}\\

\multirow{2}{*}{\shortstack{Average Clustering Coefficient}}&  $ACC(G) = \frac{1}{n} \sum_{i=1}^n c(u_i), u_i \in V, n = |V| $ & \multirow{2}{*} {~\cite{chakrabarti2007graph,li2011graph,kairam2012graphprism,chakrabarti2006graph,mislove2007measurement}}
\\ & $c(u_i) = \frac{2T(u_i)}{\binom{degree(u_i)}{2}}$, where $T(u_i)$ is the number of triangles through node $u_i$   &  \\
% \\ & $c(v) = \frac{|\{(u,w)|u,w \in \Gamma(v) , (u,w) \in E\}|}{|\Gamma(v)|(|\Gamma(v)|-  1)/2}, v,u,w \in V$  &  \\

\shortstack{Global Clustering Coefficient} %\\(transitivity)
& $GCC(G) =\frac{ 3 \times |\triangles|}{|\connectedtriples|}$& ~\cite{chakrabarti2006graph,kairam2012graphprism}  \\

Square Clustering& $SCC(G) =  \frac{1}{n} \sum_{i=1}^n c_4(u_i), u_i \in V, n = |V| $ & \multirow{2}{*} ~\cite{lind2005cycles} \\& where $c_4(u_i)$ is the quotient between the number of squares which $u_i$ participated and the total number of possible squares &\\
% Square Clustering& $SCC(G) =  \frac{1}{n} \sum_{i=1}^n c_4(u_i), u_i \in V, n = |V| $ & \multirow{2}{*} ~\cite{lind2005cycles} \\& $c_4(u_i) = \frac{\sum_{u=1}^{k_v}\sum_{w=u+1}^{k_v}q_v(u,w)}{\sum_{u=1}^{k_v}\sum_{w=u+1}^{k_v}[a_v(u,w)+q_v(u,w)] }$, where, $q_v(u,w)$ is the number of common neighbors of $u$ and $w$ other than $v$, $a_v(u,w)=(k_u - (1+q_v(u,w)+\theta_{uv}))(k_w - (1+q_v(u,w)+\theta_{uw}))$, $\theta_{uw} = 1$ if u and w are connected and 0 otherwise&\\

Average Path Length& $APL = \mean \{\frac{ \sum_{v\in V} \dist(u,v), u\ne v}{n-1}\}$&~\cite{chakrabarti2007graph,li2011graph,chakrabarti2006graph,mislove2007measurement}\\

% Degree Assortativity& $r = \frac{\sum_{jk}jk(e_{jk}-q_jq_k) }{\sigma_q^2}$, where $q_k $ is distribution of remaining degree, $\sigma_q$ is the standard deviation of $q_k$, $e_{jk}$ is the joint probability distribution of remaining degrees of the two vertices &~\cite{newman2003mixing,mislove2007measurement}\\
Degree Assortativity& $r = \frac{\sum_{xy}xy(e_{xy}-a_xb_y) }{\sigma_a \sigma_b}$&~\cite{newman2003mixing,mislove2007measurement}\\
%Power Law Distribution&$p(x)=Ax^{-~\gamma}, ~\gamma>1, x\geq x_{min}$&\cite{chakrabarti2007graph,chakrabarti2006graph,kairam2012graphprism,mcglohon2011statistical,mislove2007measurement} \\

Diameter& $\diam(G) = \max\{\dist(v,w), v,w \in V \}$ & ~\cite{chakrabarti2007graph,mcglohon2011statistical,kairam2012graphprism,mislove2007measurement}\\

Density & $\den = \frac{2|E|}{|V|(|V| - 1)}$ & \\

Ratio of Triangles & Rt $=  \frac{|\triangles|}{\binom{|V|}{3}}$ & \\

Node Connectivity & Cv: the minimum number of nodes to remove to disconnect the graph & ~\cite{even1975network}\\

Edge Connectivity & Ce: the minimum number of edges to remove to disconnect the graph & ~\cite{even1975network}\\

%Average Effective Eccentricity&$diamA(G) = ave\{e(u)|u \in V \}$ &~\cite{li2011graph,chakrabarti2006graph}\\

%Maximum Effective Eccentricity\\(effective diameter) &$diamM(G) = max\{e(u)|u \in V \}$ &~\cite{chakrabarti2007graph,kairam2012graphprism,mcglohon2011statistical,chakrabarti2006graph}\\

%Minimum effective eccentricity\\(effective radius)&  $rad(G) = min\{e(u)|u \in V \}$ &~\cite{li2011graph,chakrabarti2006graph}\\

%Characteristic path length& $CPL = median\{\frac{n-1}{ \sum_{v\in V} d(u,v), u\ne v}\}$&~\cite{mcglohon2011statistical,chakrabarti2006graph}\\

%Hop&$ N_h =  \sum_uN_h(u)$ &~\cite{chakrabarti2007graph,chakrabarti2006graph,kairam2012graphprism,mislove2007measurement}\\

%Percentage of central points&$\frac{|\{u \in V : rad(G) = e(u)\}|}{|V|}$ &~\cite{li2011graph}\\

%Spectral radius&$\rho(G) = |\lambda_1|$&~\cite{li2011graph}\\

 %\\ [1ex] adds vertical space
\hline %inserts single line
\end{tabular}}
\vspace{-0.5cm}
\end{table*}

%\smallskip\noindent{\bf Graph Properties:}  
\subsection{Graph Properties}
\label{sec:graph_stats}
Graph mining is applied in different domains from bioinformatics and chemistry, to software engineering and social science. Essential to graph mining is the efficient calculation of various graph properties \ross{(e.g., diameter, density)} and \ross{summary} statistics \ross{(e.g., averages, modes)} that can provide useful insight about the structural properties of a graph. \ross{For consistency, we refer to the graph properties and summary statistics as \emph{graph properties}}. 
We reviewed recent graph mining systems and identified some of the most frequently extracted \vahan{graph properties}. 
% A review of recent graph mining systems  identified some of the most frequently extracted statistics.
We list those, along with their definitions, in Table~\ref{Table:properties}. These properties range from basic, e.g., vertex count and  edge count, to complex, e.g., clustering coefficients~\cite{chakrabarti2007graph,li2011graph,kairam2012graphprism,chakrabarti2006graph,mislove2007measurement} and average path length~\cite{chakrabarti2007graph,li2011graph,chakrabarti2006graph,mislove2007measurement}. Many of them can be used to derive further \vahan{graph properties}. 
For example, graph density can be determined directly as the ratio of the number of edges $|E|$ to the maximum number of edges possible 
$|V|\times (|V| - 1)/2$, 
and real-world graphs are often found to have a low graph density~\cite{melancon2006just}. Node connectivity and edge connectivity may be used to describe the resilience of a graphs~\cite{cartwright1956structural,loguinov2003graph}, 
and graph diameter~\cite{hanneman2005introduction} captures the maximum among all pairs of shortest paths~\cite{albert1999internet,broder2000graph}. 

Other graph properties measure how tightly nodes are grouped in a graph. For example,  clustering coefficients have been used to describe many real-world graphs and can be measured locally and globally. Nodes in a highly connected clique tend to have a high local clustering coefficient, and a graph with clear clustering patterns will have a high global clustering coefficient~\cite{feld1981focused,frank1982cluster,karlberg1997testing,newman2003structure}.
Studies have shown that the global clustering coefficient
has been found to typically be larger in real-world graphs than in Erd\"os-R\'enyi graphs with the same number of vertices and edges~\cite{chakrabarti2006graph,newman2003structure,uzzi2005collaboration}, and a small-world graph should have a relatively large average clustering coefficient~\cite{davis2003small,ebel2002scale,watts1998collective}. 
Small-world graphs also have an average path length (APL) that is logarithmic in the number of vertices, while real-world graphs have small (often constant) APL~\cite{davis2003small,ebel2002scale,newman2003structure,uzzi2005collaboration,van2004yeast,watts1998collective}. 

Degree distribution is one frequently used {\color{black}property} describing the graph degree statistics. Many real-world graphs, including communication, citation, biological and social graphs, have been found to follow a power-law shaped degree distribution
\cite{boccaletti2006complex,chakrabarti2006graph,newman2003structure}. Other real world graphs have been found to follow an exponential degree distribution \cite{guimera2003self,sen2003small,wei2009worldwide}.
Degree assortativity is of particular interest in the study of social graphs and is  
calculated based on the Pearson correlation between the vertex degrees of connected pairs~\cite{newman2002assortative}. A random graph generated by Erd\"os-R\'enyi model has an expected assortative coefficient of $0$.   
Newman~\cite{newman2002assortative} extensively studied assortativity in real-world graphs and found that social networks often have positive assortativity, i.e., vertices with a similar degree preferentially connect together, whereas technological and biological graphs tend to have negative assortativity implying that vertices with a smaller degree tend to connect to high degree vertices. Assortativity has been shown to affect clustering~\cite{maslov2004detection}, resilience~\cite{newman2002assortative}, and epidemic-spread~\cite{boguna2002epidemic}.

Note that there are many other graph properties, but many of them are local, defined on the level of individual vertices, or individual edges. Examples include degree centrality of a vertex, betweenness centrality of a vertex or an edge, etc. Since we are interested in global properties, properties of the graph as a whole, we cannot directly use such local properties. We represent graphs by the $10$ properties in Table~\ref{Table:properties}, i.e., we represent each graph as a single data point in $\mathbb{R}^{10}$.
%10$-dimensional Euclidean space.

%Given the diversity of properties that are used to characterize graphs, it is critical to understand this issues of ``same stats, different graphs.'' Visualization can be key to providing insights in how graph properties are changing. However, understanding how changes in these properties reflect in graph structures and the degree in which structures can vary ...

%\smallskip\noindent{\bf Graph Generators:}
\subsection{Graph Generators}
\label{sec:graph_generators}
 Graph properties have been used to describe various classes of graphs (e.g., geometric, small-world, scale-free) and a variety of algorithms have been developed to automatically generate graphs that mimic these various properties.
Charkabati et al.~\cite{chakrabarti2007graph} divide graph models and generators into the following broad categories:
\begin{compactenum}
\item Random Graph Models: The graphs are generated by a random process.
\item Preferential Attachment Models: In these models, the ``rich get richer," as the graph grows, leading to power law effects. 
%\item Optimization-Based Models: Here, power laws are shown to evolve when risks are minimized using limited resources. 
\item Geographical Models: These models consider the effects of geography (i.e., the positions of the nodes) on the topology of the graph. This is relevant for modeling router or power grid graphs.
\end{compactenum}

The Erd{\"o}s-R\'enyi (ER) graph model is a simple graph generation model~\cite{chakrabarti2006graph} that creates graphs either by choosing a graph randomly with equal probability from a set of all possible graphs of size $|V|$ with $|E|$ edges~\cite{gilbert1959random} or by creating each possible edge of a graph with $|V|$ vertices with a given probability $p$~\cite{erdos1959random}. The latter process gives a binomial degree distribution that can be approximated with a Poisson distribution. 
Note that fixing the number of nodes and using $p = 1/2$ results in a uniform sampling from the space of isomorphic graphs. 
Even though this graph generator does not sample uniformly from the space of non-isomorphic graphs, the probability of having two isomorphic graphs if we generate $k$ graphs with $|V| = n$ according to ER graph generator with $p = 1/2$ is greater than or equal to
\begin{equation}
\label{eq:prob_calc_iso}
1 - {k\choose 2}*\frac{n!}{2^{n(n-1)/2}}.
\end{equation}
Table~\ref{Table:finding_isomorphic_probability} shows the number of graphs for fixed number of vertices $|V|$ that one needs to generate to have a high probability of avoiding two isomorphic graphs in the sample.

\begin{table*}[]
\caption{This table demonstrates the relationship between the graph nodes $|V| = n$, the size of the sample $k$ and the probability of having two isomorphic graphs in the sample calculated by \vahan{the formula \eqref{eq:prob_calc_iso} defined in Sec.~\ref{sec:graph_generators}}.}
\label{Table:finding_isomorphic_probability}
\begin{tabular}{|l|l|l|l|l|l|l|l|l|l|l|l|l|}
\hline
$n$      & 5    & 6    & 7    & 8    & 9    & 10   & 11    & 12     & 13      & 14         & 15          & 16             \\ \hline
k          & 2    & 2    & 4    & 14   & 60   & 440  & 4,240 & 55,500 & 985,000 & 23,800,000 & 787,000,000 & 35,600,000,000 \\ \hline
probability & 0.88 & 0.98 & 0.99 & 0.99 & 0.99 & 0.99 & 0.99  & 0.99   & 0.99    & 0.99       & 0.99        & 0.99           \\ \hline
\end{tabular}
\vspace{-0.3cm}
\end{table*}

The Watts and Strogatz~\cite{watts1998collective} (WS) model addresses a limitation of the ER model. Specifically, the WS model can be used to generate graphs that exhibit small-world properties and that have higher clustering coefficients.  \vahan{However, the WS model can generate disconnected graphs. We utilize the variation suggested by Newman and Watts~\cite{newman1999scaling} to ensure connectivity as some graph properties are not well defined for disconnected graphs.}

It is also possible to create graphs where the degrees follow other common probability distributions, e.g., exponential~\cite{deng2011exponential} or Gaussian~\cite{javarone2015gaussian}. Graphs with any given degree sequence can be generated using the configuration model~\cite{newman2003structure}.
Models have also been proposed for generating synthetic scale-free graphs with a scaling exponent~\cite{price1976general}.

The model proposed by Gilbert~\cite{gilbert1961random}, the geometric model (GE), places nodes according to a Poisson point process in some metric space (e.g., the unit square in 2D), and adds edges between pairs of nodes that are within a pre-specified distance threshold.
Barabasi and Albert (BA)~\cite{barabasi1999emergence} described another popular model for generating undirected graphs. It is a graph growth model in which each added vertex has a fixed number of edges $|E|$, and the probability of each edge connecting to an existing vertex $v$ is proportional to the degree of $v$. Dorogovtsev et al.~\cite{dorogovtsev2000structure} and Albert and Barabasi~\cite{albert2002statistical} also developed a variation of the BA model with a tunable scaling exponent. \ross{We use the NetworkX version of the BA model in this paper, which is a direct implementation of the original BA model~\cite{barabasi1999emergence}.}    

\subsection{Generating Non-Isomorphic Graphs}
The graph generators in the previous subsection generate graphs by sampling the space of all isomorphic graphs. 
However, there is also work on generating graphs from the set of all non-isomorphic graphs with fixed number of vertices. % uniformly at random. 
McKay~\cite{mckay1998isomorph} proposes an isomorphic-free generation method, which is known as canonical deletion. Such generators rely on an algorithm (implemented in the Nauty program) \vahan{to efficiently test} whether two graphs are isomorphic~\cite{mckay1981practical}. 
% propose the original implementation of an algorithm to test whether two graphs are isomorphic or no (the software is known as Nauty). 
The algorithm has been revised to improve performance~\cite{mckay2014practical} (in terms of running time); for example, testing whether two graphs on $100$ vertices are isomorphic takes $1$ second on average. Goldberg~\cite{goldberg1992efficient} proposes several algorithms to generate non-isomorphic graphs uniformly at random. Even though the algorithm provides good theoretical guarantees on the running time and space, % generation of graphs in polynomial time by using  polynomial space, however,
it is practical only for low-order graphs.
There are other methods to generate graphs from the set of all non-isomorphic graphs uniformly at random~\cite{tinhofer1990generating,dixon1983random}, but they are also not practical beyond low-order graphs. 
%The algorithm of Tinhofer~\cite{tinhofer1990generating} is called Restarting Procedures-Wormald's Method to generate unlabeled graphs uniformly at random, where for graphs with $n$ vertices, each graph is being produced in expected time $O(n^2)$. 

%\vahan{The non-isomorphic graph generators, described above, mostly work well for graphs with smaller number of vertices. Most of them include theoretical analysis and bounds for the running time and complexity. However, for larger graphs ($|V| > 15$), we are not aware of any algorithm that would run in a reasonable time.} 

\subsection{Exploring Graph Properties}
Bach et~al.~\cite{bach2012interactive} introduce an interactive system to create random graphs that match user-specified properties based on a genetic algorithm. The properties considered are $|V|, |E|$, average vertex degree, number of components, diameter, ACC, density, and the number of clusters (as defined by Newman and Girvan~\cite{girvan2002community}). The goal is to generate graphs that get as close as possible to a set of target properties; however, there are no guarantees that the target values can be obtained. 
Somewhat differently, we are interested in creating graphs that match several target properties exactly, but differ drastically in other parameters. 

Kennedy et al.~\cite{kennedy2017graph} provide an interactive graph analysis system called Graph Landscape, which allow researchers to explore graphs, graph sets, and benchmark collections regarding their properties. Unlike our paper, the system aims to enable the analysis of differences and similarities between different sets of graphs and to assess their value for experimental evaluations. 
%Different than this project, they are aiming to include at least 20 statistics in their system and easily study a graph set by those statistics and several 2-D projections of those statistics. 

\begin{figure}[t]
\centering
\includegraphics[width = 1\columnwidth]{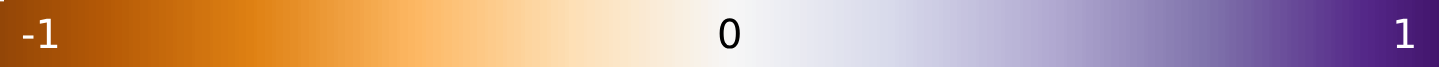}

\includegraphics[width = 0.49\columnwidth]{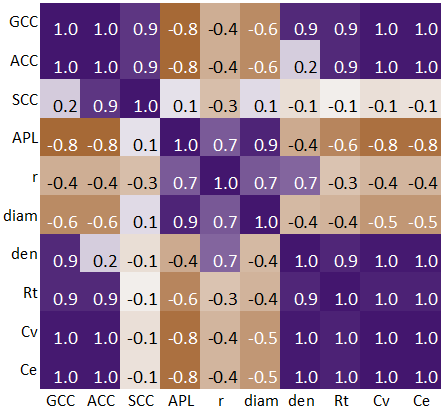}
\hfill
\includegraphics[width = 0.49\columnwidth]{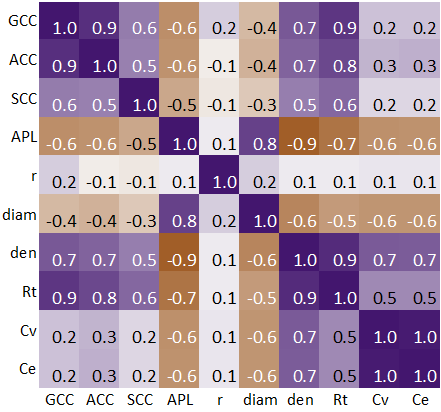}
\caption{Graph properties correlation matrix plots for the edge density dataset (left) and the ground truth set of all non-isomorphic graphs on $|V|=9$ vertices (right). 
%average clustering coefficient dataset
}
\vspace{-.5cm}
\label{Fig:heatmap}
\end{figure}

Also related is work on graph anonymization, where the goal is to generate one or more graphs with same set of fixed properties as those in a given source graph~\cite{wu2010survey, sun2013privacy, aggarwal2011hardness}. As the given graph could contain sensitive data, the generated graphs can be used instead in order to preserve anonymity. There are various kinds of graphs anonymization algorithms, each of which serve different purposes. Some examples of anonymization algorithms include K-neighborhood anonymity, edge randomization and cluster based generalization; see survey by Wu et al.~\cite{wu2010survey}. Although related, this work is different from ours as only certain parts of the graph need to be modified and only certain graph properties need to be maintained, e.g., preserving the spectral information of the underlying graph as in Ying et al.~\cite{ying2008randomizing}.
%The main application of graph anonymization is associated with the graphs that arise in social network. The data from social network contains private and sensitive information about users, so before making it available for researchers anonymization is required. A possible method for graph anonymization is to keep the graph properties that do not contain sensitive/private information and are intended for sharing and randomize the others, so it is impossible to retrieve any sensitive information. For example, Ying et al.~\cite{ying2008randomizing} proposes a spectrum preserving approach. As we have mentioned earlier, our target is to come up with a graph generator, where we can specify which statistics to keep constant and which ones to vary randomly.
\vspace{-.25cm}
\section{Preliminary Experiments and Findings}
\label{sec:preliminary_findings}
Unlike the traditional setting of Anscombe’s quartet, in the graph setting some properties are correlated with each other. Thus, fixing one property (e.g., high edge density) may allow us to vary a property such as assortatitivity, but not other properties such as diameter. This lead us to study such correlations, which in turn lead us to graph generators (as the ground truth data is too large for $|V|>10$), which in turn lead us to evaluating the generators, which in turn lead us to the qualities of the generators, which lead us to the notions of coverage and representativeness. 

In this section we present some preliminary findings about the relationships between the graph properties described in Section~\ref{sec:graph_stats}. Note that our analysis focuses on the 10 dimensional space, where each dimension corresponds to one of 10 graph properties. As discussed in Section~\ref{sec:graph_stats}, there are many other graph properties. However, many properties are local (vertex-based or edge-based). Our choice of 10 global properties is the result of analyzing the graph mining literature and selecting the most commonly studied graph properties that can also be efficiently computed. 
Similarly, as discussed in Section~\ref{sec:graph_generators}, while there are many different graph generators, we focus on 5, ensuring that we have one from each of the three major types. 
%since as mentioned in Section~\ref{sec:graph_generators}, there are 3 major types of graph generators and we picked one or two per type.
\vahan{In Section~\ref{sec:high_order} we consider the properties of graphs with 100 vertices and next, in Section~\ref{sec:low_order_analysis} we analyze properties of graphs with up to 10 vertices.}

\begin{figure*}[t]
\includegraphics[width=2\columnwidth]
% [width=\linewidth]
{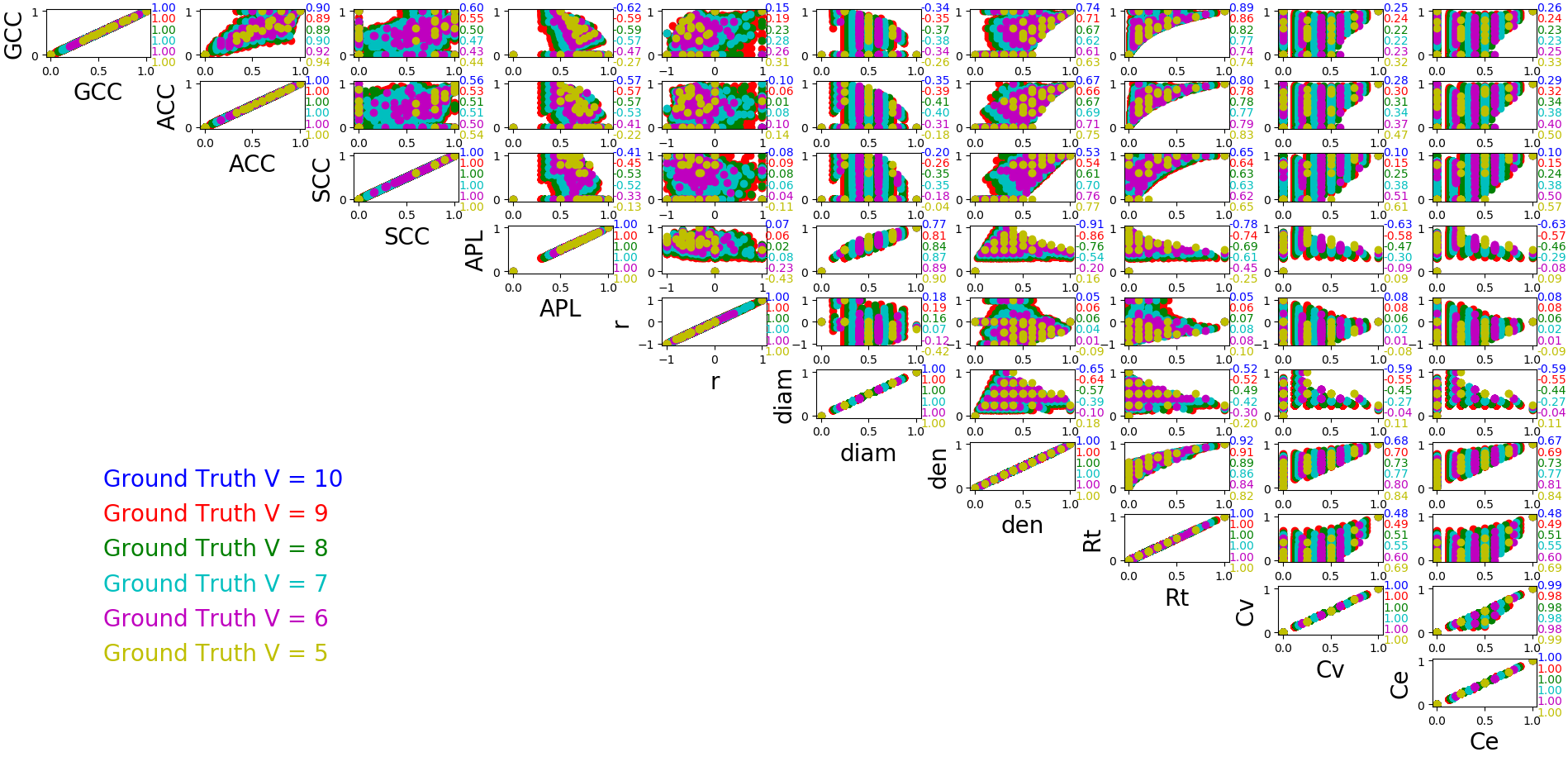}
\caption{Correlations between graph properties in the ground truth for $|V|=5,6,7,8,9$. Note that for $|V|=9$ there are already 274,668 points. Points are plotted to overlap, with the largest sets plotted first (i.e., $|V|=9, ... |V|=5$) to enable us to identify the range of properties that can be covered with a given number of vertices.}
  \label{fig:hang-style}
  \vspace{-4mm}
\end{figure*}
\subsection{Graph Properties in Higher-Order Graphs}
\label{sec:high_order}
In a recent study of the ability to perceive different graph properties, such as edge density and clustering coefficient in different types of graph layouts (e.g., force-directed, circular), we generated a large number of graphs with 100 vertices. Specifically, we generated graphs that vary in a controlled way in edge density and graphs that vary in a controlled way in the average clustering coefficient~\cite{soni2018perception}. A post-hoc analysis of this data (http://vader.lab.asu.edu/GraphAnalytics/) reveals some interesting patterns among the properties listed in Table~\ref{Table:properties}. 

The edge density dataset has 4,950 graphs and for each graph, we
compute the $10$ properties described in Table~\ref{Table:properties}.
%The average clustering coefficient dataset has 3,450 graphs. 
We then compute Pearson correlation coefficients and observe high positive (purple) correlations and negative (brown) correlations for many {\color{black}property} pairs, Fig~\ref{Fig:heatmap}.
%We observed high positive (blue) correlations and negative (yellow) correlations for many property pairs. 
For example, the average clustering coefficient is highly correlated with the global clustering coefficient and the number of triangles.
%Specifically, the average clustering coefficient dataset  shows some strong correlations, e.g., the number of triangles is strongly correlated with local and global clustering coefficients and also with the square clustering coefficient.
 %, that last of which is somewhat surprising.

These graphs were created for a very specific purpose and cover only a limited space of all graphs with $|V|=100$. 
%Two factors likely impact these correlations: the generator used, and the statistical properties that were controlled (e.g., ensuring the number of edges remained constant). 
The type of generators we used, and the way we used them (some properties were controlled), could bias the results and influence the correlations. 
%For the edge density graph set, we used a random graph generator that varies the number of edges while keeping a uniform probability of connecting two nodes.  
%
%Specifically, for to generate the average clustering coefficient graph set, we controlled graph density and degree distribution (power-law). 
%By fixing one property and varying along another, it is possible that the correlations seen only exist across certain fixed parameters. However, 
The fact that these correlations exist when some properties are fixed indicates that we can keep certain graph properties fixed while manipulating others. This motivated us to conduct the following experiments:
\begin{enumerate}
\item Generate all non-isomorphic lower order graphs ($|V|\leq 10$) and analyze the relationships between {\color{black} graph} properties. We consider this type of data as ground truth due to its completeness.
\item Use different graph generators 
%to create a sample of graphs 
and compare how well they represent the space of non-isomorphic graphs and how well they cover the range of possible values in the ground truth data.
%the graph property distributions of these outputs to the ground truth. This depicts the characteristics of different graph generators in terms of graph property coverage and helps us to think about which generator we should use in different scenarios.
\end{enumerate}
%\yafeng{In terms of the same stats, different graphs problem, these experiments could explain the possible ranges one graph properties varies in when some are fixed.}
An analysis of the set of 274,668 non-isomorphic graphs on $|V|=9$ vertices shows that the correlations are quite different than those in graphs from our edge density experiment; see Fig.~\ref{Fig:heatmap}.

\subsection{Graph Properties in Lower-Order Graphs}
\label{sec:low_order_analysis}
We start the experiment by looking at the pairwise relationships of graph properties of low-order graphs, where all non-isomorphic graphs can be enumerated. If two properties, say $s_1$ and $s_2$, are highly correlated, then fixing $s_1$ is likely to restrict the range of possible values for $s_2$. On the other hand, if $s_1$ and $s_2$ are independent, fixing $s_1$ might not impact the range of values for $s_2$, yielding the same stats ($s_1$) for different graphs ($s_2$). We first study the correlations between the properties under consideration.

% We compute all properties for all non-isomorphic graphs on $|V|=4,5, \ldots, 10$ vertices (we exclude graphs with fewer vertices as many of the properties are not well defined and there are only a handful of graphs).
We focus on the analysis of graphs with $|V| = 5, 6,  \ldots, 10$ (unless otherwise stated) and compute all properties for all non-isomorphic graphs in this range. 
One reason we do not consider $|V| < 5$ is that  many of the properties require at least 3 vertices (e.g., GCC, ACC, ). Another reason is that the number of different graphs on 1, 2, 3 and 4 vertices is very small: 1, 2, 4 and 11.
%and for $|V| = 4$ there are only $11$ non-isomorphic graphs and for our analysis we need more graphs.

We next consider the pairwise correlations between the different graph properties
and how this changes as the graph order increases, Fig~\ref{fig:hang-style}.
To compare the coverage of graph properties with different $|V|$, we scale the values of graph properties into the same range. By definition, clustering coefficients (ACC, GCC, SCC) are in the $[0,1]$ range and degree assortativity is in the $[-1, 1]$ range. We keep their values and ranges without scaling.  
%the following five statistics, 
Edge density, number of triangles, diameter and  connectivity ($C_v$ and $C_e$), are normalized into $[0,1]$ (dividing by the corresponding maximum value). 
APL, is also normalized into $[0,1]$, subject to some complications:  we compute the exact average path length to divide by in our ground truth datasets, but not when we use the generators, where we use the maximal path length encountered instead (which may not be the same as the maximum).
%.\yafeng{[Hang: explain why this is challenging.]} Stephen: i explained in the text.

\begin{figure}[tbh]
\includegraphics[width=0.95\linewidth]{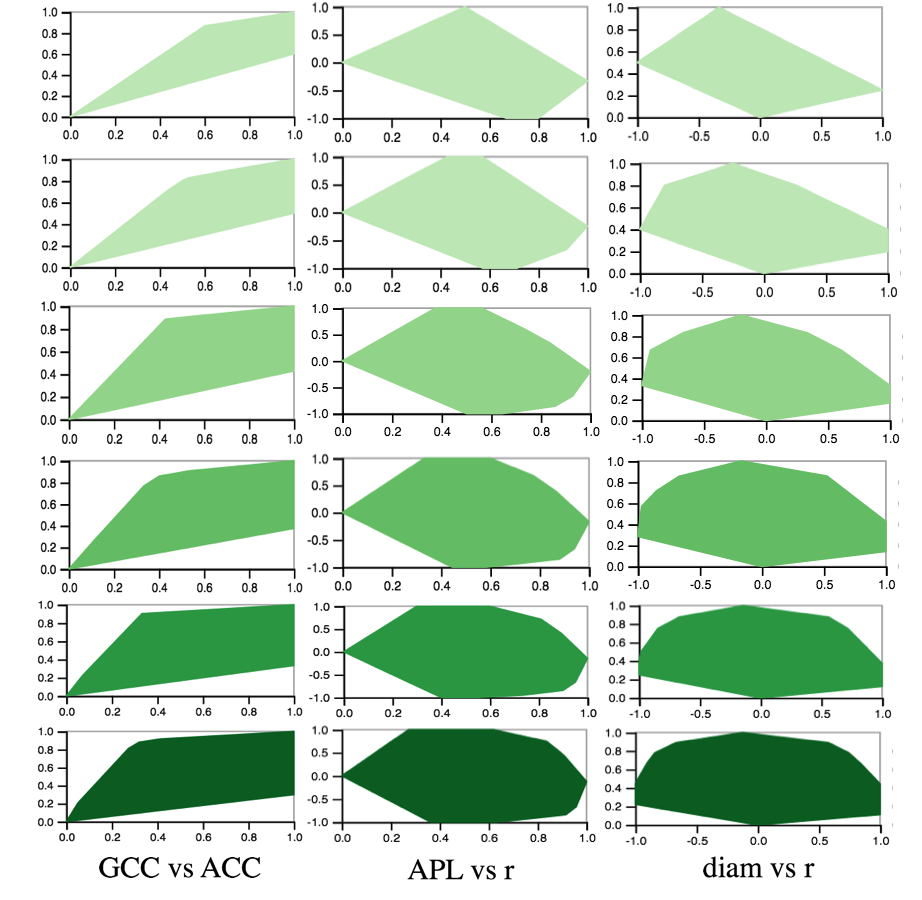}
\caption{The convex hull of graph coverage across several {\color{black}graph} properties. Each row represents all graphs for a fixed number of vertices ($|V| = 5 ... |V|=10$). Columns are pairs of graph properties. 
% The darkness of the color indicates the density (the number of points per square unit).
{\color{black}The color is uniform for each image and corresponds to the average number of graphs in the image (the more of them, the darker the color).}}
  \label{fig:coverage_comparison}
  \vspace{-2mm}
\end{figure}
\begin{figure}[h] 
\includegraphics[width=0.95\linewidth]
% [width=\linewidth]
{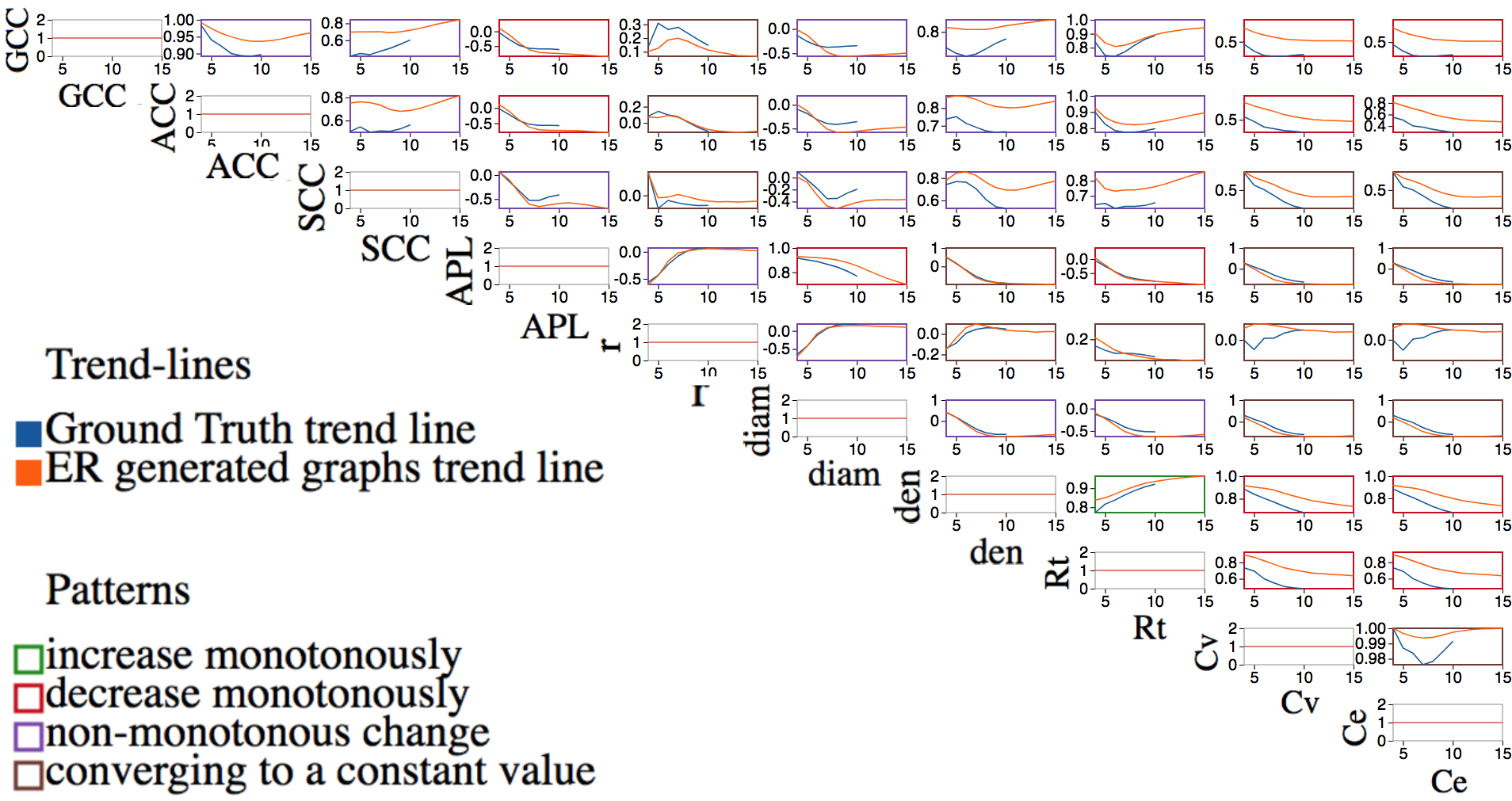}
\caption{Trends in the correlations with increasing $|V|$: the $x$-axis shows the number of vertices and the $y$-axis shows the correlation value for the pair of graph properties.}
\label{fig:correlation_vs_n}
\vspace{-5mm}
\end{figure}
It is easy to see that the coverage of values expands with increasing $|V|$. Fig.~\ref{fig:coverage_comparison} shows this pattern for three pairs of properties. 
%This indicates that we are more likely to find larger ranges of different statistics (when a set of other statistics are fixed) for graphs with more vertices.
This indicates that 
we are more likely to find larger ranges of different graph properties for graphs with more vertices given the same set of fixed properties.
With this in mind, we consider graphs with more than $10$ vertices, but this time relying on random graph generators. 
\vahan{Fig.}~\ref{fig:correlation_vs_n} shows how correlation values between all pairs of graph properties change when the number of vertices increases. The blue trend lines for the ground truth data show the correlation values calculated using the set of all possible graphs for a given number of nodes. The orange
trend lines show the correlation values calculated from graphs generated with the ER model. Specifically, the ER-model data is created as follows: for each value of $|V|=5,6,\ldots, 15$ we generate $100,000$ graphs with $p$ selected uniformly at random in the $[0,1]$ range.

For most of the cells in the matrix, Fig.~\ref{fig:correlation_vs_n}, the correlation values seem to converge as $|V|$ becomes larger than $8$ 
(both in the ground truth and the ER-model generated graph sets). 
%\yafeng{This happens for both the ground truth data and the ER-model generated data.}
The pattern of the change in correlation values appears to be the same for both sets. Analyzing the trend lines of the ER-model, we observe four patterns of change in the correlation values: convergence to a constant value, monotonic decrease, monotonic increase, and non-monotonic change.
%(also annotated in Figure \ref{fig:correlation_vs_n} by an enclosing colored box)
\vahan{Fig.}~\ref{fig:correlation_vs_n} highlights these patterns using box outlines of different colors.
There are exceptions that do not fit these patterns, e.g., ($S_c$, r) and in two cases, (r, $C_v$) and (r, $C_e$), the trend lines show different patterns.

\begin{figure}[tbh]
\includegraphics[width=\columnwidth,height=2.3cm]{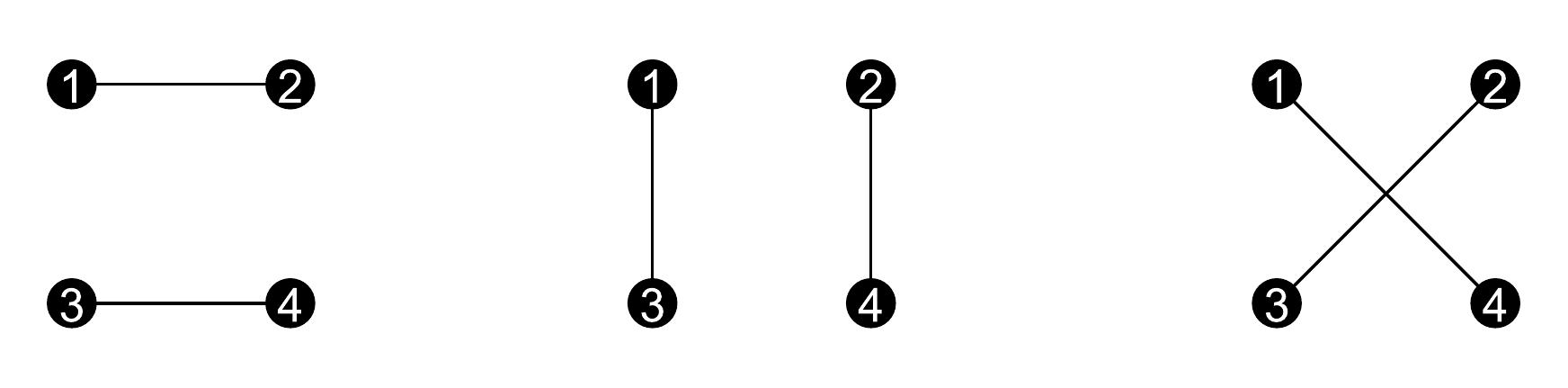}
\caption{All isomorphic graphs $|V| = 4, |E| = 2$ whose edges are disjoint.}
\label{fig:disjoint}
\end{figure}
\begin{figure}
% \centering
% \includegraphics[width=\columnwidth,height=2.3cm]{./images/2line}
\includegraphics[width=\columnwidth,height=2cm]{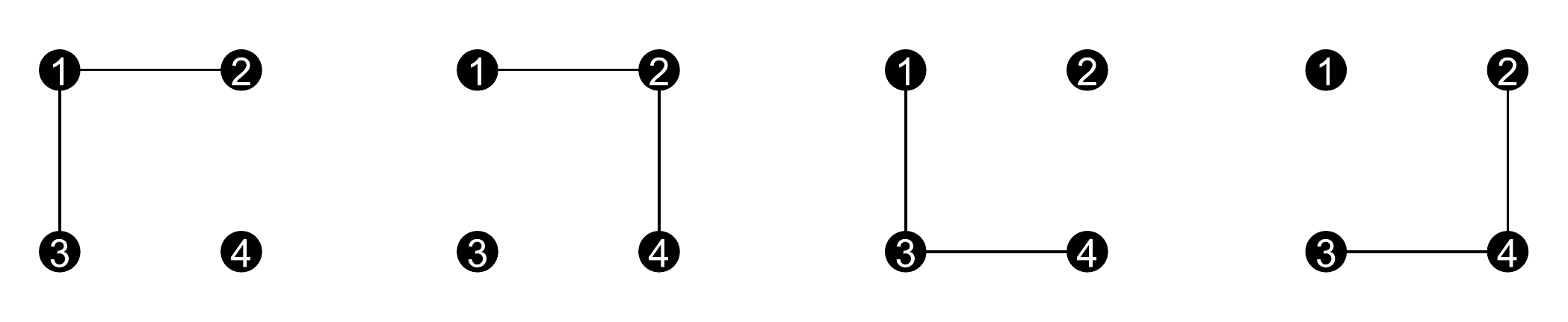}
\includegraphics[width=\columnwidth,height=2cm]{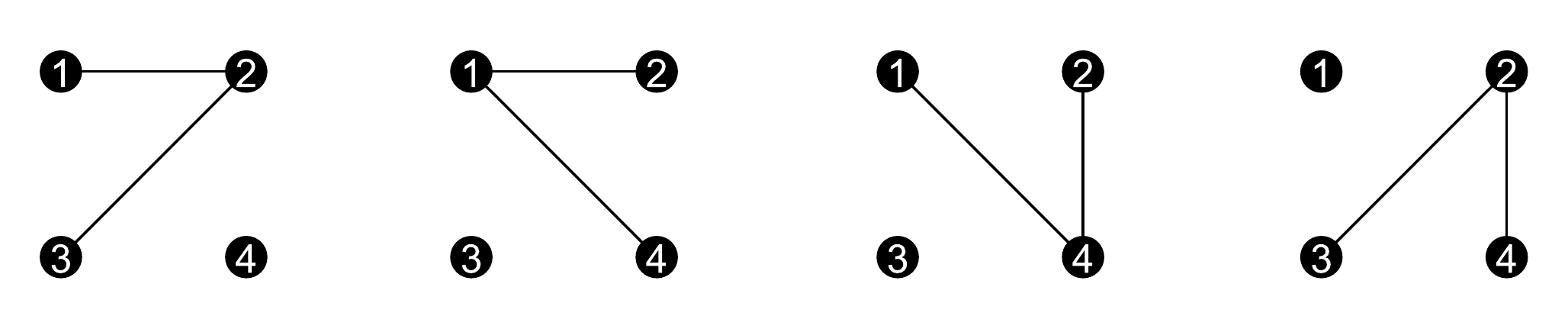}
\includegraphics[width=\columnwidth,height=2cm]{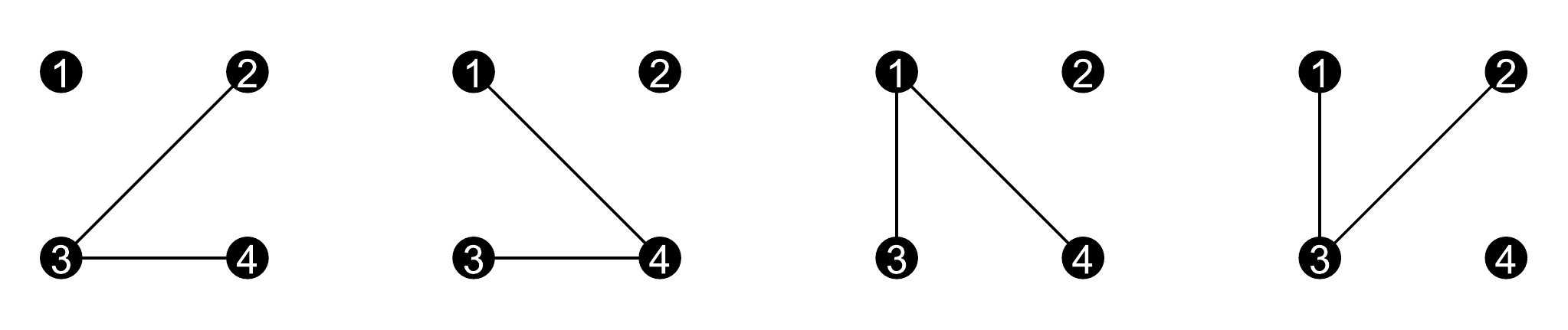}
\caption{All isomorphic graphs $|V| = 4, |E| = 2$ whose edges form a path.}
\label{fig:path}
\end{figure}
\section{Finding Same Stats, Different Graphs}
\label{sec:finding_same_stat}
In this section we consider examples of the ``same stats, different graphs" phenomenon. \hang{In order to study and find ``same stats, different graphs," we run several small experiments:}
\begin{itemize}

\item \hang{The first experiment indicates that very sparse or very dense graphs have few isomorphic copies, while graphs with edge density close to 0.5 have many such copies}.

\item \hang{The second experiment shows that graphs with the same edge densities do not necessarily have the same number of isomorphic copies.}

\item \hang{The third experiment shows that even the best graph generator (in terms of representativeness), ER,  does not represent the space of all non-isomorphic graphs well.}
\end{itemize}

\hang{We start with the case when the two graphs are indeed isomorphic. 
Structurally the path graph 1-2-3-4 is indistinguishable from the path graph 1-3-2-4, however, for an isomorphic graph generator, these two graphs are different. 
Intuitively, very sparse or very dense graphs have few isomorphic copies, while graphs with edge density 0.5 have many such copies. 
Consider for example graphs with 4 vertices. There is only one graph for with 4 vertices and 0 edges (the empty graph) and only one graph with 4 vertices and 6 edges (the complete graph). Relabeling those graphs does not create different isomorphic copies. On the other hand, there are three structurally different graphs with 4 vertices and with 3 edges; their degree sequences are \{2, 2, 1, 1\}, \{3, 1, 1, 1\}, \{2, 2, 2, 0\}. Since their degree sequences are different, changing labels leads to different isomorphic copies.}

%\hang{Warning, this is not true for all cases, for example circle have degree = 2 for all node and circle does not have such copies.}

\begin{table*}[tbh]
\centering
\caption{{\color{black} Illustration of the number of repetitions in the set of $12,346$ graphs generated by the ER model with $|V| = 8$ and $p = 1/2$. The size of the sample is equal to the number of non-isomorphic graphs with $|V| = 8$. The table indicates the number of sets of graphs having the same exact $10$ properties (although they may not be isomorphic). For example, there are 2 sets with 8 graphs that each have the exact same 10 properties. To put this in perspective, in the ground truth there are only 7 pairs of ``same 10 stats, but different graphs,'' while in this sample we have $1,955$ such pairs.}}
\begin{tabular}{|l|l|l|l|l|l|l|l|l|}
\hline
\# of sets with graphs that have the same $10$ properties                   & 2 & 7 & 37 & 109 & 348 & 833 & 1955 & 3713 \\ \hline
\# of graphs in the set      & 8 & 7 & 6  & 5   & 4   & 3    & 2    & 1    \\ \hline
\end{tabular}
\label{table:repeatitions_ER_8}
\end{table*}

It is then natural to ask whether two graphs with the same edge densities have the same number of isomorphic copies. Unfortunately, this is not the case, as illustrated in Fig.~\ref{fig:disjoint} and Fig.~\ref{fig:path} by two small graphs with $|V| = 4$ and $|E|=2$. Structurally, there are only two different types of graphs with 4 vertices and 2 edges. In the first type the the two edges form a path; see Fig.~\ref{fig:path}. In the second type the two edges are disjoint; see Fig.\ref{fig:disjoint}. Importantly, there are only 3 graphs for which the edges are disjoint, while there are 12 graphs for which the edges form a path.
If a pair of graphs is isomorphic then they have the same properties, including the 10 properties that we are measuring. However, a pair of graphs can be structurally different (non-isomorphic) while still having the same properties.
To find such low-order graphs we can explore the entire set of non-isomorphic graphs (as we have explicit representations for all of them). 
For larger graphs, we use the graph generators together with some filters.

\begin{figure}[ht!]
\begin{center}
\includegraphics
[width=0.25\columnwidth,angle=90]
{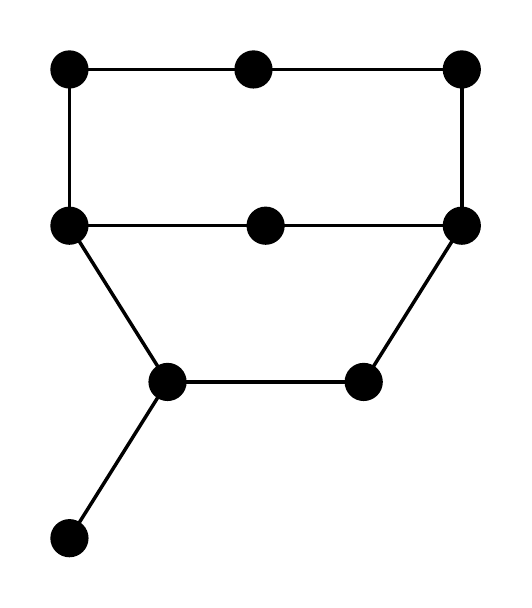}
\hspace{-0.3cm}
\includegraphics[width=0.29\columnwidth]{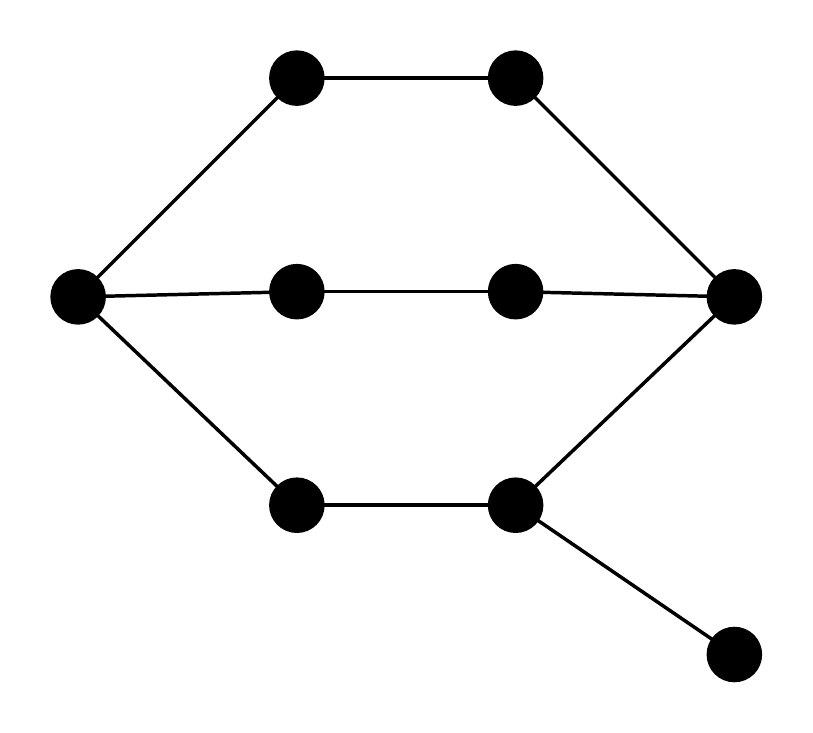}
\includegraphics[width=0.29\columnwidth]{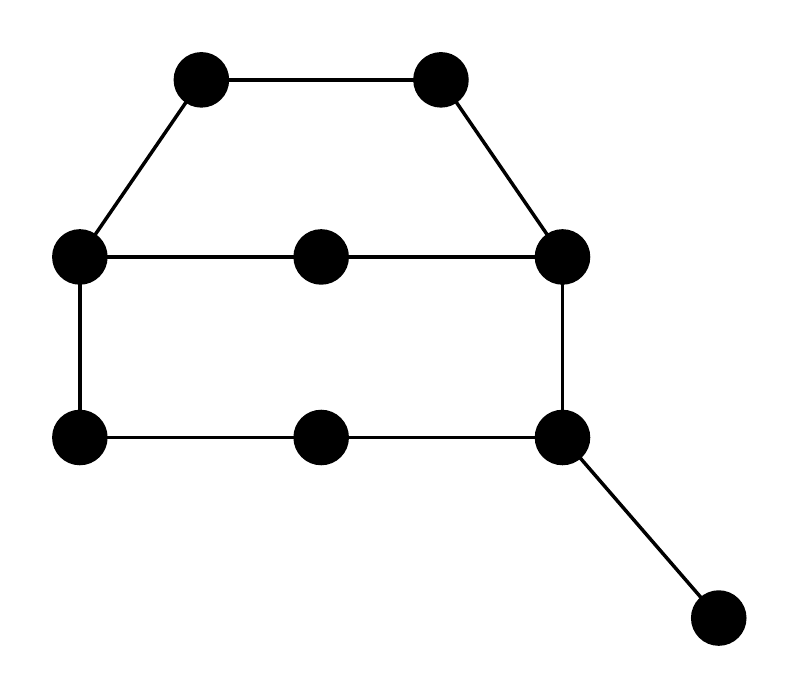}
\end{center}
\caption{An example of three different graphs that have exactly the same $10$ properties: $9$ verties, $10$ edges, $GCC=ACC=SCC=Rt=0$, $APL=2.11$, $r=-0.47$, $diam=4$, $Cv=Ce=1$}
\label{fig:same_stats_graphs}
\end{figure}

%\hang{the following text is repeat in conclution}
Examining the ground truth data for $|V| = 7$, we find  a pair of graphs that have exactly the same 10 properties. For $|V| = 8$, we find $8$ pairs of graphs that have exactly the same 10 properties. For $|V|>8$ we have found many more graphs with exactly the same 10 properties, such as the triple of graphs shown in Fig.~\ref{fig:same_stats_graphs}.

%As we mentioned earlier, we are unaware of any non-isomorphic graph generators. 
In Sec.~\ref{sec:rep_results} and Sec.~\ref{sec:cov_results} we analyze the 5 graph generators in terms of {\em representation} and {\em coverage}, and we explore their performance as generators of non-isomorphic graphs. We use the ER generator with probability of adding any edge $p = 1/2$ and $|V| = 8$, and generate 12,346 graphs (equal to the number of non-isomorphic graphs on $|V|=8)$). We compute the number of repetitions (over the $10$ properties). %discussed in Sec.~\ref{sec:graph_stats}). 
Results are given in Table~\ref{table:repeatitions_ER_8}.
% {\color{red} discuss what this means!}

\begin{figure}[ht!]
\begin{center}
\includegraphics[width=0.34\columnwidth]
% [width=0.44\columnwidth,height=2.3cm]
{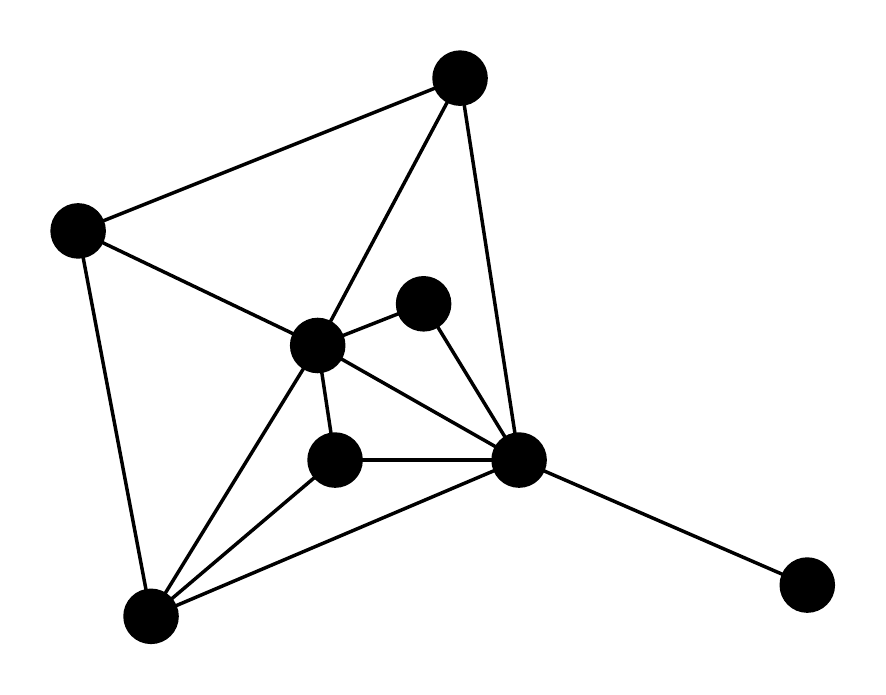}
\includegraphics[width=0.34\columnwidth]
% [width=0.44\columnwidth,height=2.3cm]
{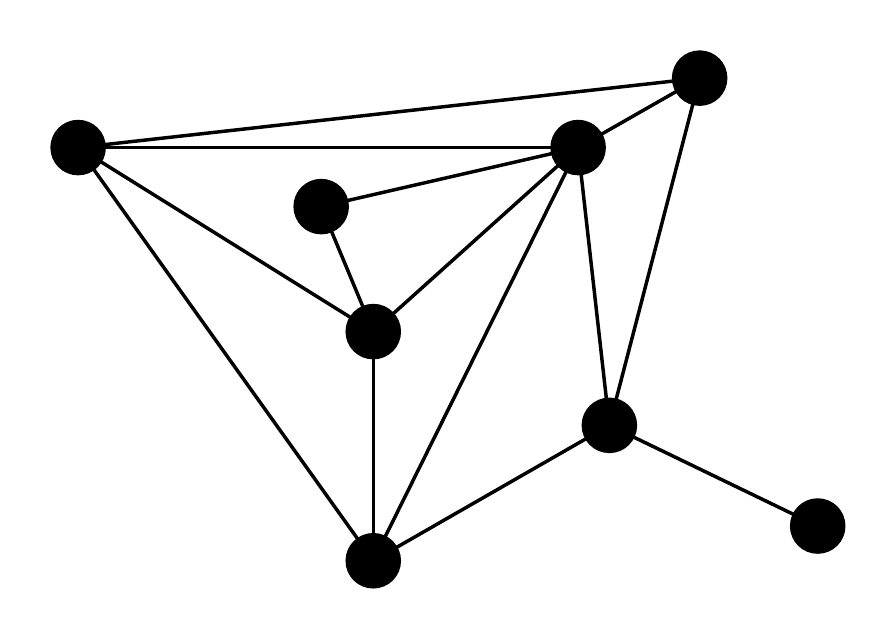}
\end{center}
\caption{Illustration of the graphs that appeared most frequently in the sample generated by the ER model with $p = 1/2$, $V = 8$. Each of the two graphs appeared $8$ times.} 
\label{fig:high_freq_graphs}
\end{figure}

Comparing to the total set of non-isomorphic graphs for $|V|=8$, the set generated by ER has more graphs with the same properties. 
% The ER model generated $7,004$ distinct graphs (in terms of graph properties) out of the $12,346$ graphs in the ground truth set. 
There are 1,955 pairs that have exactly the same 10 {\color{black}graph properties} in the sample, while in the total set of non-isomorphic graphs there are only 7 such pairs. %there are $12346$ non-isomorphic graphs, but only $7$ pairs of graphs that have exactly the same $10$ properties. 
\vahan{Fig.}~\ref{fig:high_freq_graphs} shows the two most common graphs that appear in the ER model samples.

%\smallskip\noindent{\bf Finding Graphs in the Ground Truth:} 
\subsection{\hang{A Tool for Finding Similar Graphs}}
We provide an analytic tool (\url{http://findgraph.cs.arizona.edu/}) for exploring the ``same stats, different graphs'' phenomenon, that is, looking for graphs with several fixed properties and one that varies. In this tool, we separate our task into two parts and integrate them in one interface. For graphs with $|V| \leq 10$ we examine all possible non-isomorphic graphs. For graphs with $|V|> 10$ we use graph generators and a filter.
%We discuss these details in sections \ref{sec:finding_graphs} and \ref{sec:finding_graphs_generator}.

\hang{For graphs with $|V|\leq 10$, since the ground truth is known, we know whether graphs with fixed several properties exist or no. In order to find such graphs, we create one interactive parallel coordinate plot (PCP) for each set of non-isomorphic graphs with fixed $|V|$. We provide an online version of PCP (\url{http://differentgraphs.cs.arizona.edu/pcp/index.html}). In order to make the waiting time affordable, we use k-means clustering instead of all non-isomorphic graphs with fixing $|V|$.}

\subsection{Finding Similar Low-Order Graphs}
\label{sec:finding_graphs}
For $|V|\leq 10$, we examine our dataset of all possible non-isomorphic graphs, looking for graphs with several fixed properties and one that varies.  We use a spring layout to visualize this series of same stats, different graphs.
%that have encapsulate the variability of one statistic in 10 slots, covering the ranges $[0.0, 0.1], [0.1, 0.2], \dots [0.9,1]$ and in each slot we show a graph (if it exists) drawn by a spring layout; see Fig.~\ref{fig:assort}.

\begin{figure*}[ht]
\includegraphics[width=2\columnwidth]{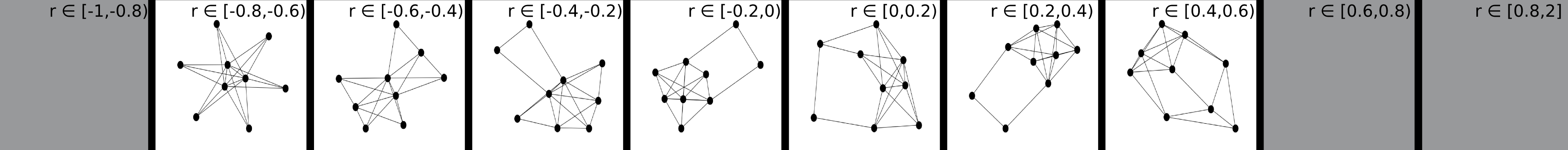}
\caption{{\color{black}Examples of graphs with fixed number of vertices $|V| = 9$ and $4$ other fixed properties ($APL\in (1.42,1.47)$, $den \in (0.52,0.57)$, $GCC\in (0.5,0.6)$, $R_t\in (0.15,0.25)$), but with varying assortativity values (between -1 and 1 by jumps of 0.2). The empty slots represent that there is no graph with the specified set of properties.}}  
\label{fig:assort}
%\label{fig:gcc_vary},
%\label{fig:Ce}
\end{figure*}

For the first experiment, we fix $|V| = 9$, $APL\in (1.42,1.47)$, $den\in (0.52,0.57)$, $GCC\in$ (0.5,0.6), $R_t\in (0.15,0.25)$. 
Since all our properties are normalized to $[0,1]$ and assortativity is in $[-1, 1]$, we consider 10 possible ranges for assortativity, each of size 0.2 and we find graphs for seven of the ten possible ranges; \vahan{see} Fig.~\ref{fig:assort}. This figure also illustrates the output of our ``same stats, different graphs" generator: fix several graph properties and generate graphs that vary in another properties.

\begin{figure*}[ht]
\includegraphics[width=2\columnwidth]{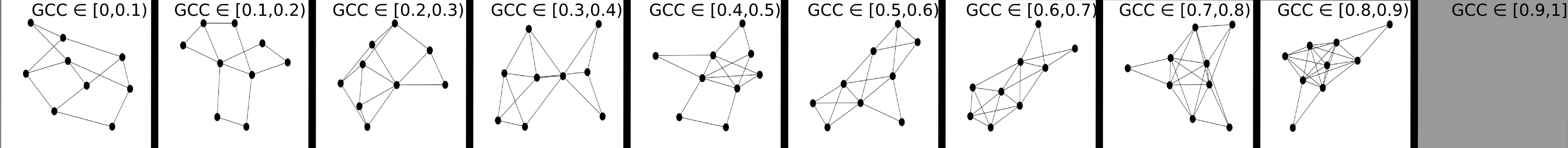}
\caption{{\color{black}Examples of graphs with fixed number of vertices $|V| = 9$ and $4$ other fixed properties ($\diam = 3$, $Cv = 2$, $Ce = 2$, and $r \in ( -0.29, -0.22)$), but with varying global clustering coefficient (GCC in the range between 0 and 1 by jumps of 0.1). The empty slots represent that there is no graph with the specified set of properties.}}
\label{fig:gcc_vary}
\end{figure*}

\begin{figure*}[h!]
\hspace{.3cm}
\includegraphics[width=2\columnwidth]{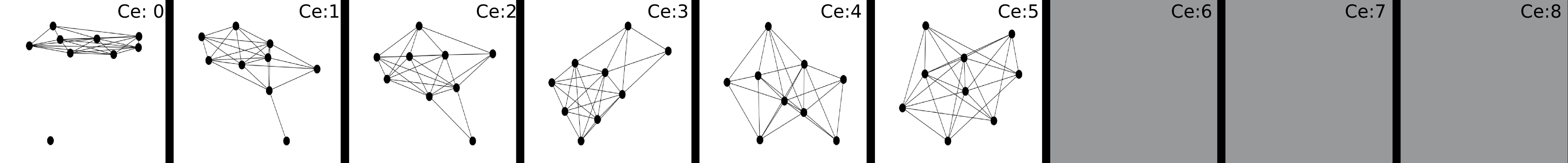}
\caption{{\color{black}Examples of graphs with fixed number of vertices $|V| = 9$ and $4$ other fixed properties ($SCC\in (0.75,0.85)$, $ACC\in (0.75,0.8)$, $r\in (-0.3, -0.2)$, $R_t\in (0.35,0.45)$), but with varying edge connectivity (in the range between 0 and 8 by jumps of 1). The empty slots represent that there is no graph with the specified set of properties.}} 
\label{fig:Ce}
\vspace{-0.4cm}
\end{figure*}

Similarly, for the second experiment, we fix $|V| = 9$, $\diam = 3$, $Cv = 2$, $Ce = 2$, and $r \in ( -0.29, -0.22)$ and look for graphs with different $GCC$ values. GCC is in the range $0$ to $1$ and we find graphs for nine of the ten possible ranges {\color{black}by jumps of 0.1; Fig.~\ref{fig:gcc_vary}}.

As a final example, we fix $|V| = 9$, $SCC\in (0.75,0.85)$, $ACC\in (0.75,0.8)$, $r\in (-0.3, -0.2)$, $R_t\in (0.35,0.45)$ and find graphs with varying connectivity. Specifically, we look for different values of edge connectivity, $C_e$, which is in the range 0 to $|V|-1$; Fig.~\ref{fig:Ce}.

Note that the graphs in Figures~\ref{fig:assort}-\ref{fig:Ce} are different in structure even though they possess similar values for many properties. For example, in Fig.\ref{fig:gcc_vary}, there are more and more triangles as we go from left to right.

%\smallskip\noindent{\bf Finding Graphs Using Graph Generators:} 
\subsection{Finding Similar Graphs Using Generators}
\label{sec:finding_graphs_generator}
This approach relies on generating many graphs and filtering graphs based on several fixed properties. 
For the two most important properties of a graph, $|V|$ and $|E|$, we generate all graphs with a fixed $|V|$ and choose $|E|$ as follows:
%Ross: These two \items are confusing to me. Not sure how to rewrite.
\begin{enumerate}
\item UN: select $|E|$ uniformly from its range. This is equivalent to forcing the edge density in the generated set to follow a uniform 
distribution;

\item ER: select $|E|$ from by the binomial distribution, that is, each edge appears with fifty percent probability.

% \item population: select $|E|$ by forcing the edge density in the generated set to match the distribution in the ground truth (population) graph set.
\end{enumerate}

\begin{figure}[tbh]
\includegraphics[width=\linewidth]{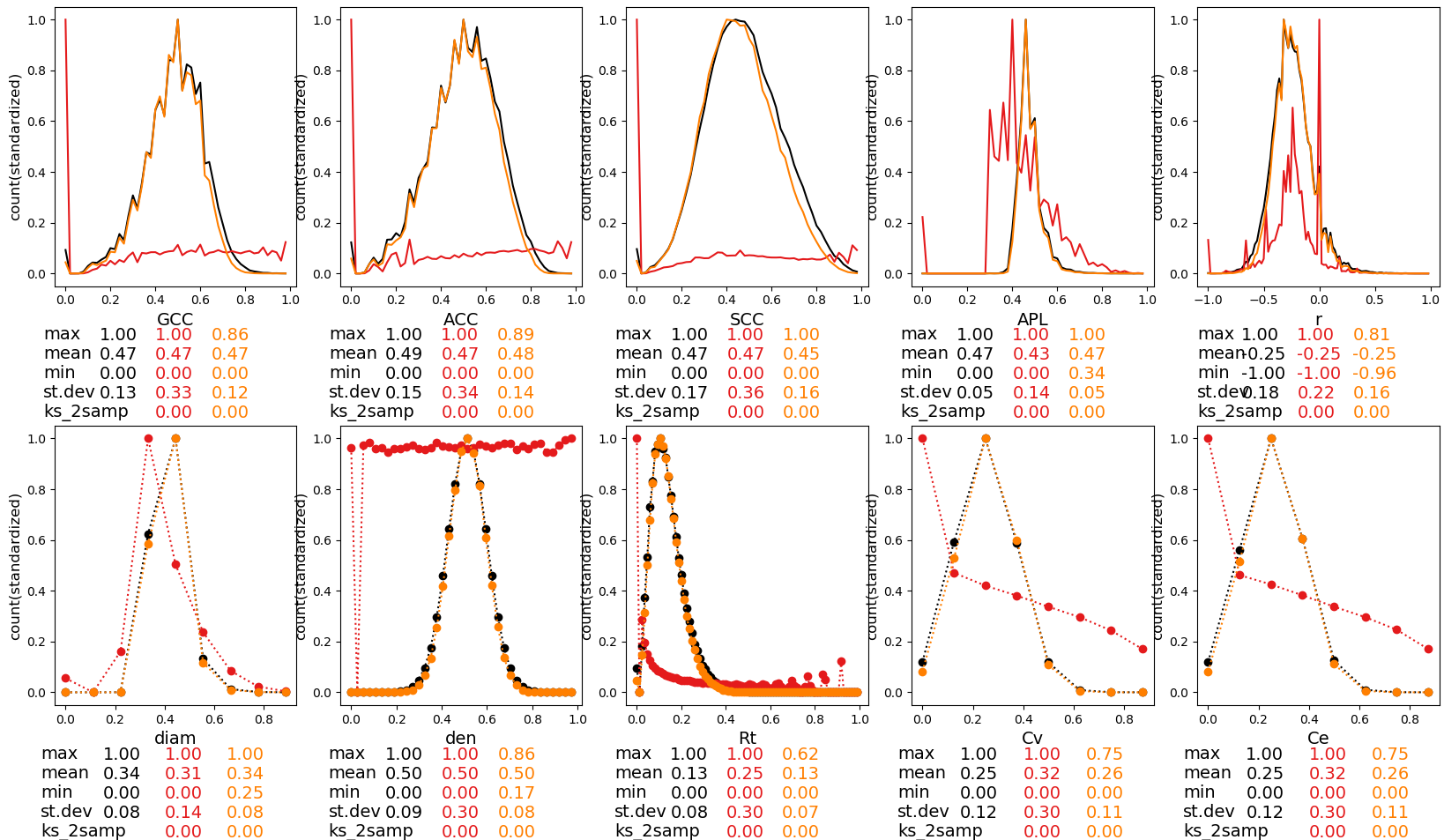}
\caption{Distribution of the $10$ properties, including min/mean/max and standard deviation. Ground truth is colored black, ER with p = 1/2 in orange, UN in red.}
\label{fig:sdd} 
\vspace{-.4cm}
\end{figure}

Using both edge selection strategies for all four generators, we compare the statistics distribution to the ground truth for $|V|=9.$ \vahan{Fig.}~\ref{fig:sdd} illustrates how different {\color{black}properties} are distributed for the UN and ER generators. It shows that although the ER generates a distribution that is more similar to the ground truth, it does not cover the range of values (larger min and smaller max) than the UN generator. The WS and BA models also do not provide good coverage of the various graph properties. We next discuss the notions of representation and coverage of the ground truth.
%see the Appendix for more details. 

\section{Generator Evaluation}
\label{sec:rep_cov}
For low order graphs (in this setting graphs with $|V| \le 10$), we can explore statistical coverage and representation of a generated sample by comparing it with the set of all graphs with fixed number of vertices.
However, it is difficult to generate all non-isomorphic graphs with more than $10$ vertices due to the super-exponential increase in the number of graphs (e.g., for $|V|=16$ there are $6\times 10^{22}$ different graphs). Nevertheless, these higher order graphs are common in many domains. We turn to graph generators in order to further explore the issue of ``same stats, different graphs" for larger graphs.

In this section, we discuss two approaches for measuring the quality of the statistical approximation of the set of properties for the sampled graphs when compared against the ground truth. Note that we are going to generate graphs with a fixed number of vertices (e.g., $|V|=9$) and compare the sample to the set of all different graphs with the same number of vertices. The comparison is done with respect to the $10$ properties defined in Sec.~\ref{sec:graph_stats}. {\color{black}The first measure, which we refer to as \textit{representation}, evaluates the extent to which the set of sampled graphs represents the properties of the set of all graphs with fixed number of vertices. The second measure, which we refer to as \textit{coverage}, evaluates the extent to which the sampled set of graphs covers a similar range of values as the set of all graphs with a fixed number of vertices. Both settings refer to the $10$ dimensional space defined in Sec.~\ref{sec:graph_stats}.}
 
We analyze graphs sampled from the following five models: ER with probability $1/2$ (ER), ER with $p$ selected uniformly at random from the  $[0,1]$ range (UN), geometric (GE), Watts and Strogatz (WS), and Barabasi and Albert (BA). We use the implementations of the  generators (ER, WS, BA, GE) from NetworkX~\cite{hagberg2008exploring}.

\begin{figure}[tbh]
\centering
 \includegraphics[width=0.8\columnwidth]{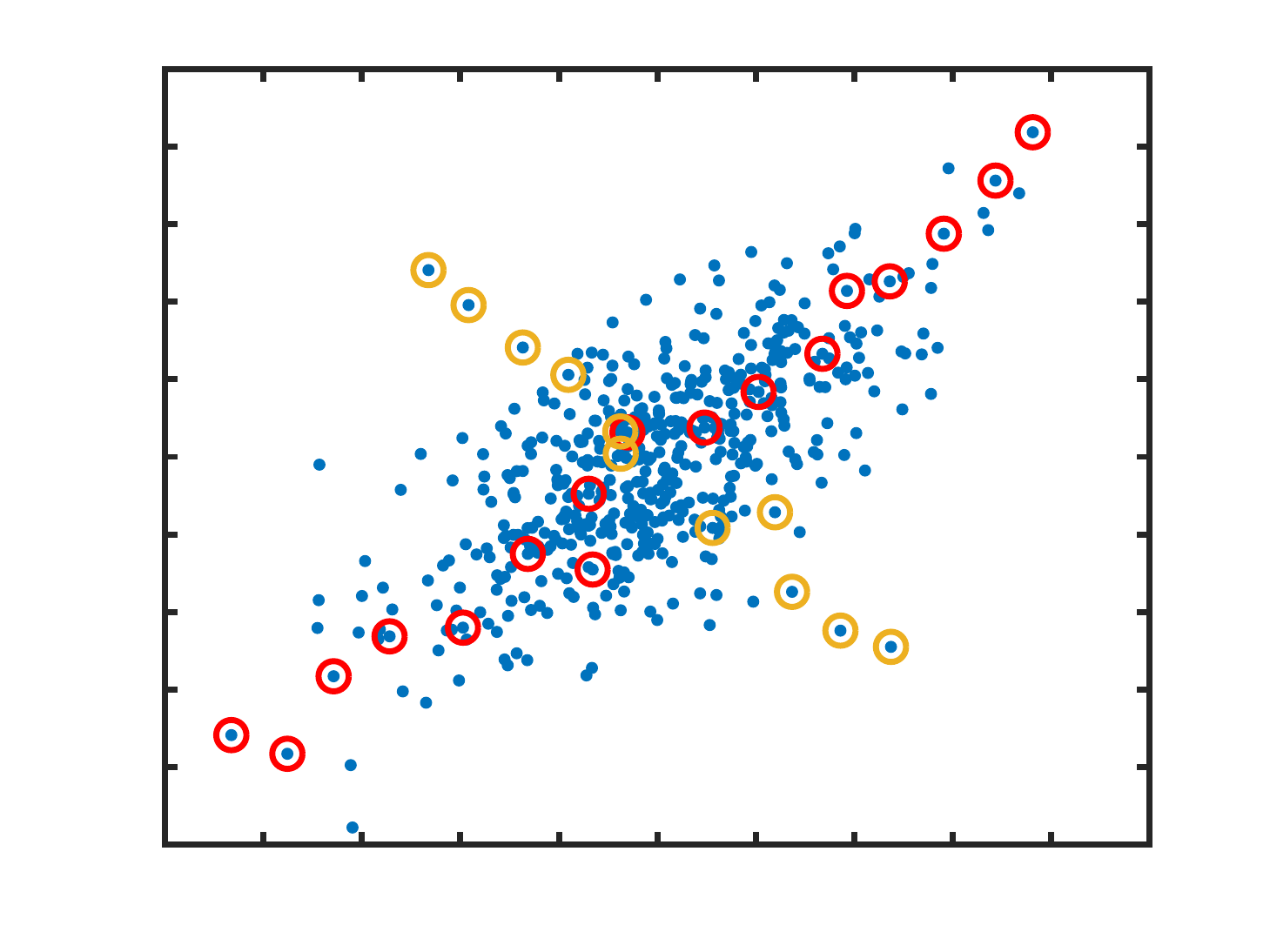}
 \caption{Example of a dataset in $2$D where the diameter of the sample (points circled in red) is the same as the diameter for the ground truth (all points in blue), but the sample does not cover the dataset well (see points with yellow circle).}
 \label{fig:diam_limit}
 \end{figure}

\subsection{Representation}
\label{sec:representation}
Our goal here is to explore whether a small sample of graphs with a fixed number of vertices can represent the set of all non-isomorphic graphs with the same number of vertices and how this representation changes as the sample size becomes larger (i.e., going from 1\% to 100\%). For this purpose, we review and analyze the following four methods: Pearson correlations, Kolmogorov-Smirnov (KS) test, Kullback-Leibler (KL) divergence, and Wasserstein distance, which is also known as the earth mover (EM) distance. 

\begin{figure*}[tbh]
\includegraphics[width=2\columnwidth]{./images/colorbar}
\includegraphics[width=0.66\columnwidth]{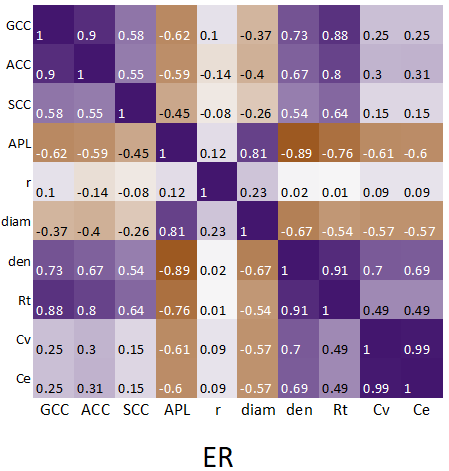}
\includegraphics[width=0.66\columnwidth]{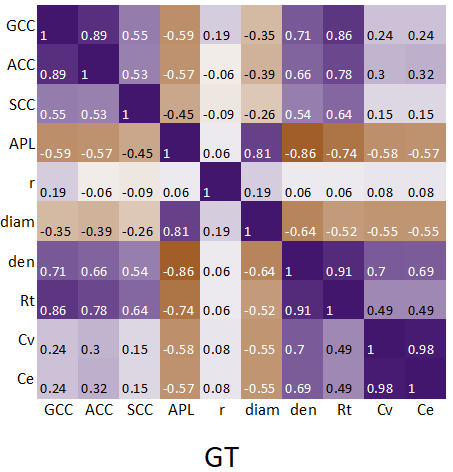}
\includegraphics[width=0.66\columnwidth]{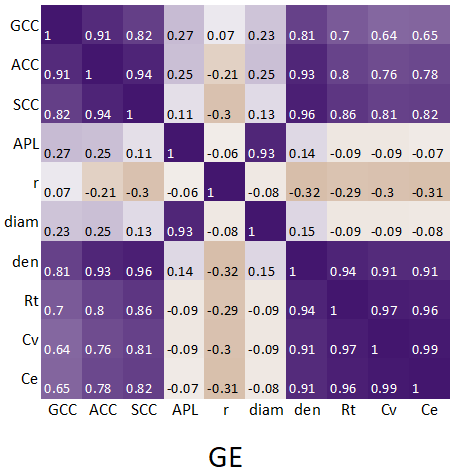}
\caption{Relative correlations between the $10$ properties for graphs with $|V| = 9$. The left table shows the correlations for the dataset sampled by the ER $p = 1/2$ graph generator of size $1\%$ of the ground truth dataset. The middle table shows correlations for the ground truth. The right table presents the relative correlations for the dataset sampled by the GE graph generator of size $1\%$ of the ground truth dataset.}
\label{fig:correlations}
\end{figure*}

One possible way to measure how representative a graph generator is to generate graphs with it, compute the graph properties described in Table~\ref{Table:properties}, calculate relative correlations between the graph properties, and compare the results with those in the ground truth. Since we consider $10$ properties we have $45$ such comparisons, Fig.~\ref{fig:correlations}, which makes it difficult to compare the different generators.

The KS test~\cite{massey1951kolmogorov} 
is a nonparametric test used to compare a sample with a reference distribution (one-sample case), that is, to quantify how well the sample represents the given distribution. The KS test can also be used to compare two samples (two-sample case) and quantify whether both samples represent the same distribution. The null distribution of this statistic is calculated under the null hypothesis that the samples are drawn from the same distribution (in the two-sample case). In our setting, we need to compare sampled data with the ground truth. To do so, we propose to uniformly sample $10\%$ of the ground truth and use the KS test with the generated sample and the uniformly sampled dataset. We repeat this procedure 10 times and average the results. However, similar to pairwise correlations, the KS test results in $10$ different numbers, one for each {\color{black}property}. 

%To overcome the setbacks of relative correlations and KS test, we suggest to use KL divergence and the EM distance. 
Unlike relative correlations and the KS test, KL divergence and EM distance would give us a single value associated with the generated dataset. However, since we have a discrete dataset and the underlying distribution is unknown, we use the formulas for a multinomial normal distribution. For this, one needs to calculate the mean and the covariance of the generated data and the ground truth, then use the formulas for KL divergence and EM distance:
\small
\begin{equation*}
    \begin{split}
    D_{KL} (N_0 | N_1) = \frac{1}{2} \biggr{(} \tr(\Sigma_1^{-1} \Sigma_0) + (\mu_0 - \mu_1)^T \Sigma_1 (\mu_0 - \mu_1)  \\
    - k + \ln \left(\frac{\det(\Sigma_1)}{\det(\Sigma_0)} \right)  \biggr{)},
    \end{split}
\end{equation*}
\normalsize

\begin{equation*}
    D_{EM} (N_0 | N_1) = \Vert \mu_0 - \mu_1 \Vert_2^2 + \tr \left(\Sigma_0 + \Sigma_1 - 2(\Sigma_1^{1/2} \Sigma_0 \Sigma_1^{1/2})^{1/2} \right).
\end{equation*}

\begin{figure}[tbh]
\centering
 \includegraphics[width=0.8\columnwidth]{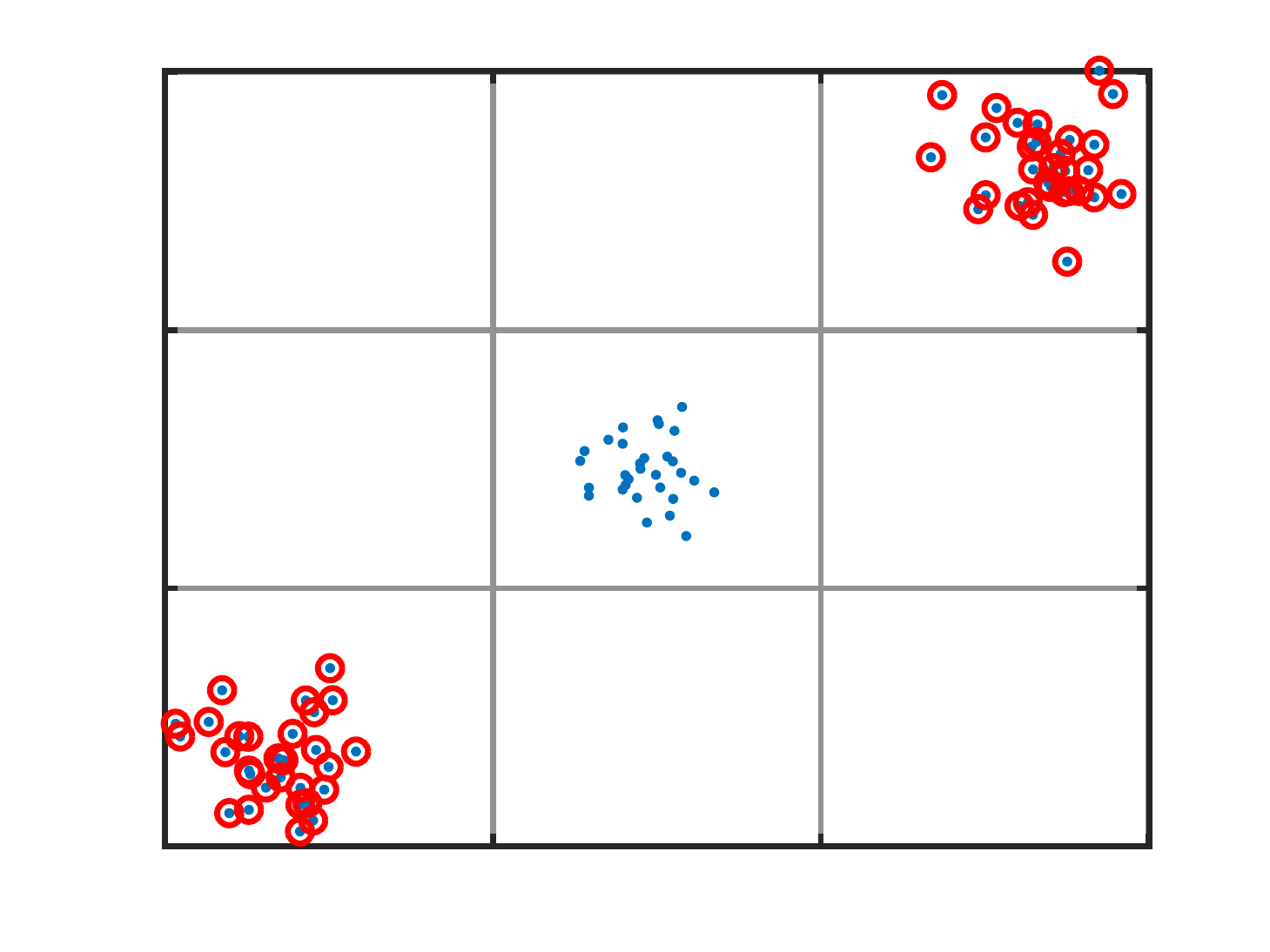}
 \caption{Example of a dataset in $2$D where the volume of the bounding box (points circled in red) is the same as the volume of the bounding box of the ground truth (all points in blue), but the sample does not cover the dataset well (see the blue points in the middle). This example demonstrates that split bounding box would perform better in this case.}
  \label{fig:sp_bb}
\end{figure}

\subsection{Coverage}
\label{sec:coverage}
In this section we introduce four different measures: diameter, volume of the bounding box, split bounding box, and volume of the robust ellipse to measure the coverage of the sampled dataset. The ultimate goal is to see whether these coverage measures are consistent for different generators and under different sample sizes. Next, we define the four measures mentioned above and discuss their advantages and limitations for our setting. 
For a discrete dataset $S \subset \mathbb{R}^{d}$, the \textit{diameter} measures the largest possible distance between all the pairs of points in $S$. The precise mathematical definition of the diameter is:
$$
\diam(S) = \sup_{x, y \in S} \Vert x - y \Vert.
$$
If the diameter of the sampled dataset is smaller than the diameter of the ground truth, then the sampled dataset does not cover the complete range of graph properties. However, if the diameter of the sampled dataset is the same, or in the same range as the diameter of the ground truth, it does not necessarily imply that the sampled dataset covers the range of the graph properties of all graphs; see Fig.~\ref{fig:diam_limit}.

The complexity of calculating the diameter of a discrete set $S$ is $O(|S|^2)$, this is the complexity of finding the distances between all possible points of $|S|$. For large datasets it would be too time consuming to compute the diameter. To overcome this issue, we uniformly subsample points from the dataset and find the diameter of the subsampled dataset. To make sure that the results are accurate, we calculate the diameter for $10$ such subsamples and report their average.

We also propose to use the \textit{volume of the bounding box}. For a set $S = \{x_1, \dots, x_n\} \subset \mathbb{R}^{d}$ the bounding box of $S$ is defined as
\begin{multline*}
BB(S) = \{ (a_i, b_i), i = 1, 2 \dots d \ | a_i = \min \{x_1(i), \dots x_n(i)\}, \\ b_i = \max \{x_1(i), \dots x_n(i) \}, \text{ for } i = 1, \dots, d \}.
\end{multline*}
The volume of the bounding box for a set $S \subset \mathbb{R}^d$ is the volume of the $d$-dimensional hyperrectangle $BB(S)$ which is $\Pi_{i=1}^d (b_i - a_i)$.

\begin{figure*} [tbh]
\centering
%  \hspace{-1cm}
 \includegraphics[width=1.0\columnwidth]{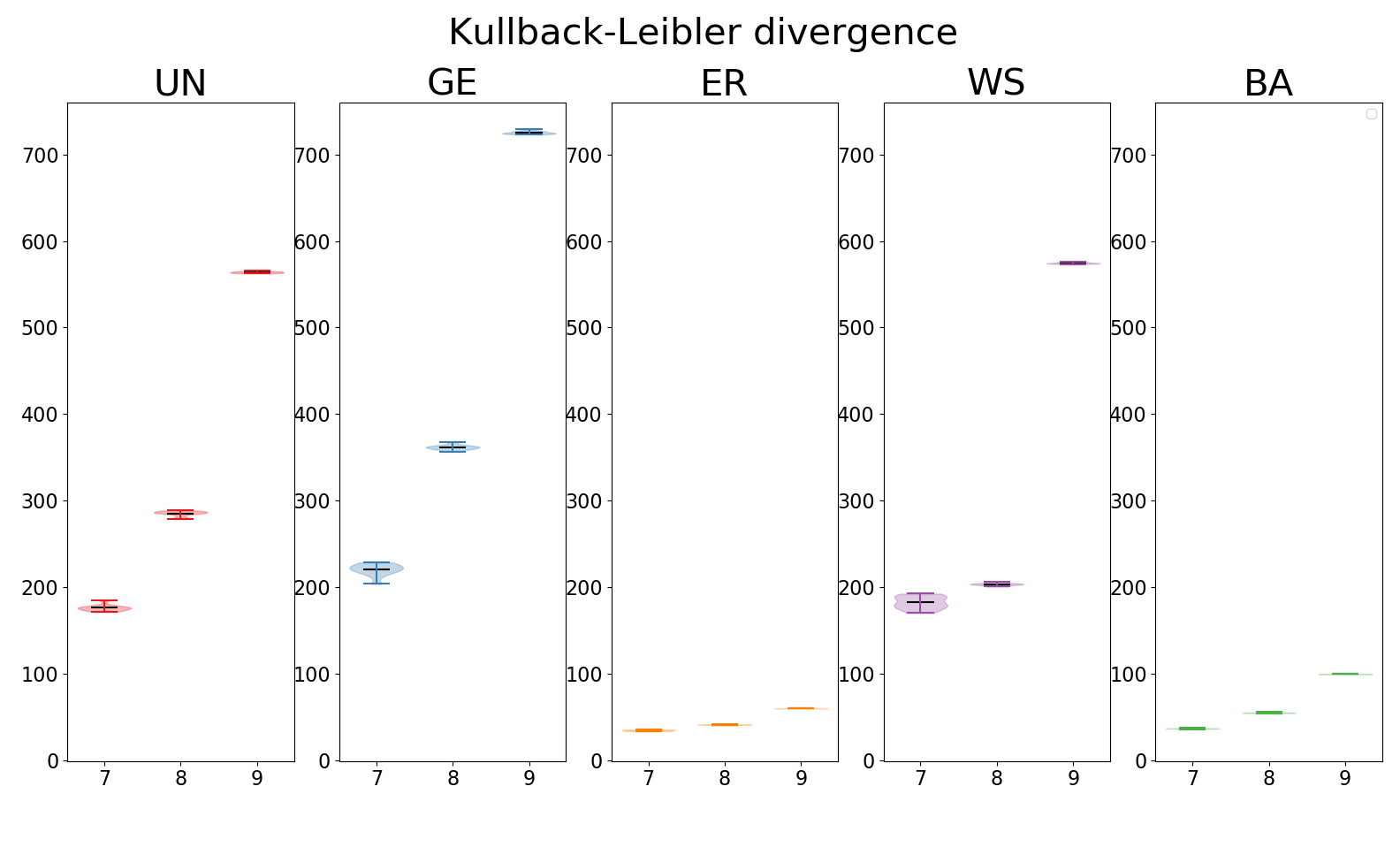}
%  \hspace{-2cm}
 \includegraphics[width=1.0\columnwidth]{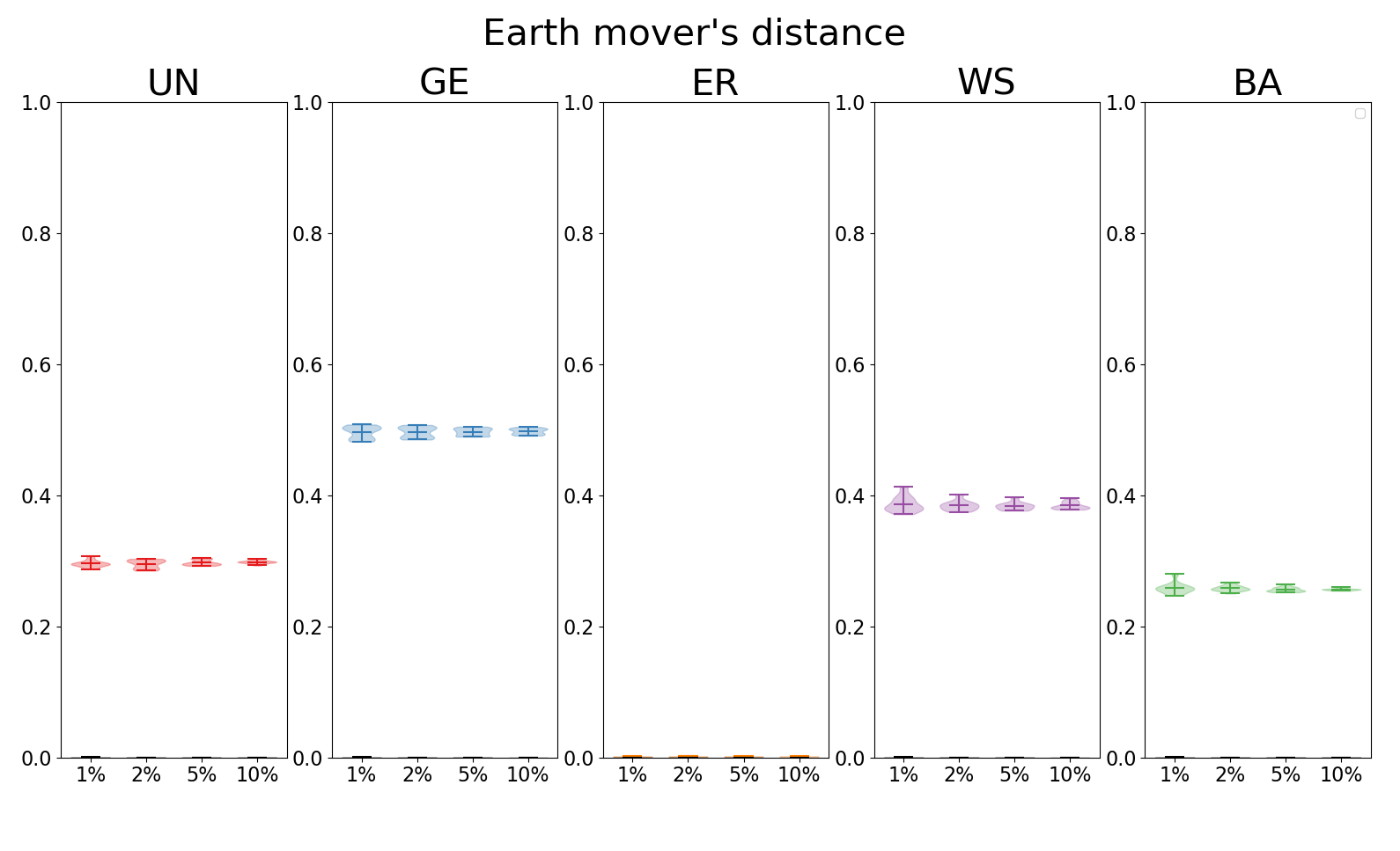}
 \caption{Representation comparison of the five graph generators (UN, GE, ER, WE and BA)  over ten samples, with sample sizes equal to the size of the ground truth dataset for $|V| = 7$, $|V| = 8$, and $|V| = 9$. KL divergence is on the left and EM distance is on the right; see Sec.~\ref{sec:representation}.}
 \label{fig:rep_measure_vertex}
 \vspace{-0.5cm}
 \end{figure*}

 \begin{table}[tbh]
\centering
\caption{The ratio of the volumes of convex hulls for sampled ($1\%$) and the ground truth in 8D (excluding Ce and Cv), $|V|$ = 9.}
\begin{tabular}{|l|l|l|l|l|l|}
\hline
                                    & UN      & GE      & ER     & WS     & BA     \\ \hline
Ratio & 11.96\% & 14.78\% & 0.81\% & 1.25\% & 0.11\% \\ \hline
\end{tabular}
\label{table:convex_hull}
\end{table}
Although the bounding box captures information for each dimension and maintains extreme point information for each dimension, similar to the diameter, it can suffer if the dataset is not concentrated around a hyperrectangle. The bounding box volume is highly influenced by outliers, especially for small sample sizes. If the ratio of the bounding boxes volumes is small it implies that the sample does not cover the ground truth; however, it is hard to make a conclusion if the ratio is around $1$; Fig.~\ref{fig:sp_bb}.

We also consider the \textit{split bounding box} measure as a generalization of the standard bounding box, where we create multiple bounding boxes that span the data. We divide the range of each data dimension into multiple parts of equal size resulting in multiple hypercubes. 
Next, for each hypercube, we compute the bounding box of the sample and the ground truth data (restricted to this hypercube), calculate the ratio of their volumes, and average them across all hypercubes; see Fig.~\ref{fig:sp_bb}. Dividing the range of each measure into $k$ equal parts results in $k^{10}$ hypercubes for our dataset. As a result, when the number of dimensions is high, this measure requires a large sample size. 
%Note that we can not afford to have a large value of $k$, as we want each hypercube to contain reasonable amount of datapoints. Such analysis is presented in Table~\ref{table:uniform_test_k}. We uniformly sampled $1,000,000$ data points in $[0, 1]^{10}$ hypercube and analyzed the behaviour of the split bounding box measure for sample sizes $10,000$, $100,000$, $200,000$ and $500,000$. As we can see in Table~\ref{table:uniform_test_k} having $k = 2$ and large enough sample provides sufficiently good results.\stephen{why use a different dataset?}
%\begin{table}[ht]
%\begin{tabular}{|l|l|l|l|l|}
%\hline
%k\textbackslash{}percentage & 1\%                  & 10\%               & 20\%                & 50\%              \\ \hline
%1               & 0.998    & 1.000  & 1.000  & 1.000 \\ \hline
%2               & 0.359    & 0.912  & 0.959  & 0.989 \\ \hline
%3               & 1.248e-05 & 0.015 & 0.094 & 0.499 \\ \hline
%4               & 1.093e-06 & 0.001 & 0.006 &  0.114 \\ \hline
%\end{tabular}
%\caption{Comparison of the split bounding box values for four different sample sizes and %different values of $k$.}
%\label{table:uniform_test_k}
%\end{table}

A possible way to overcome the limitations associated with the bounding box and split bounding box is to consider the \textit{convex hull}. We can compute the convex hull of the sample and the ground truth dataset and calculate the ratio of their volumes. However, the $O(n^{\lfloor d/2 \rfloor})$ convex hull computation is computationally expensive in high dimensions. 
%Multiple days and a high performance computing cluster were not sufficient to complete the computation for our datasets up to $|V|=10$, 
%using the QHull algorithm~\cite{barber1996quickhull}. 
%In our experiments we were not able to calculate the volume of the convex hull for a $10$ dimensional datasets. 
We show the convex hull results based on $8$ of the $10$ properties (excluding Ce and Cv) computed using the QHull algorithm~\cite{barber1996quickhull}; see Table~\ref{table:convex_hull}. 
%of their discrete structure); see Table.~\ref{table:convex_hull}. 

A  computationally efficient alternative to the convex hull is the \textit{robust ellipse} measure.
%Another idea of measuring the coverage of the generated sample is to fit an ellipse to the sample and to the total dataset, calculate their volumes and consider their ratio. However, this would have the same limitations as the diameter, the volume of the bounding box and the volume of the convex hull. Thus, we suggest a slightly different procedure. 
\begin{figure*} [tbh]
\centering
 \includegraphics[width=\columnwidth]{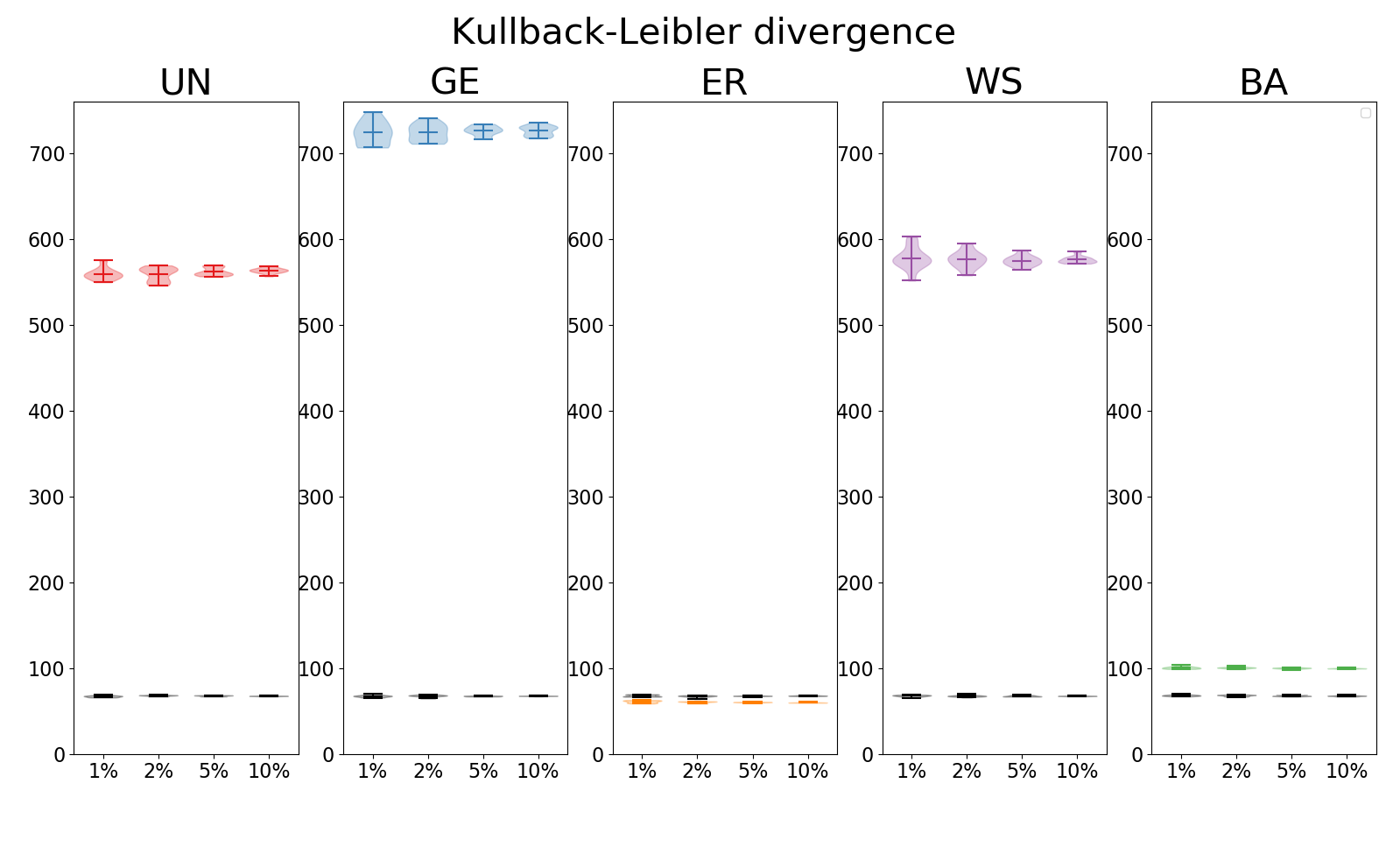}
 \includegraphics[width=\columnwidth]{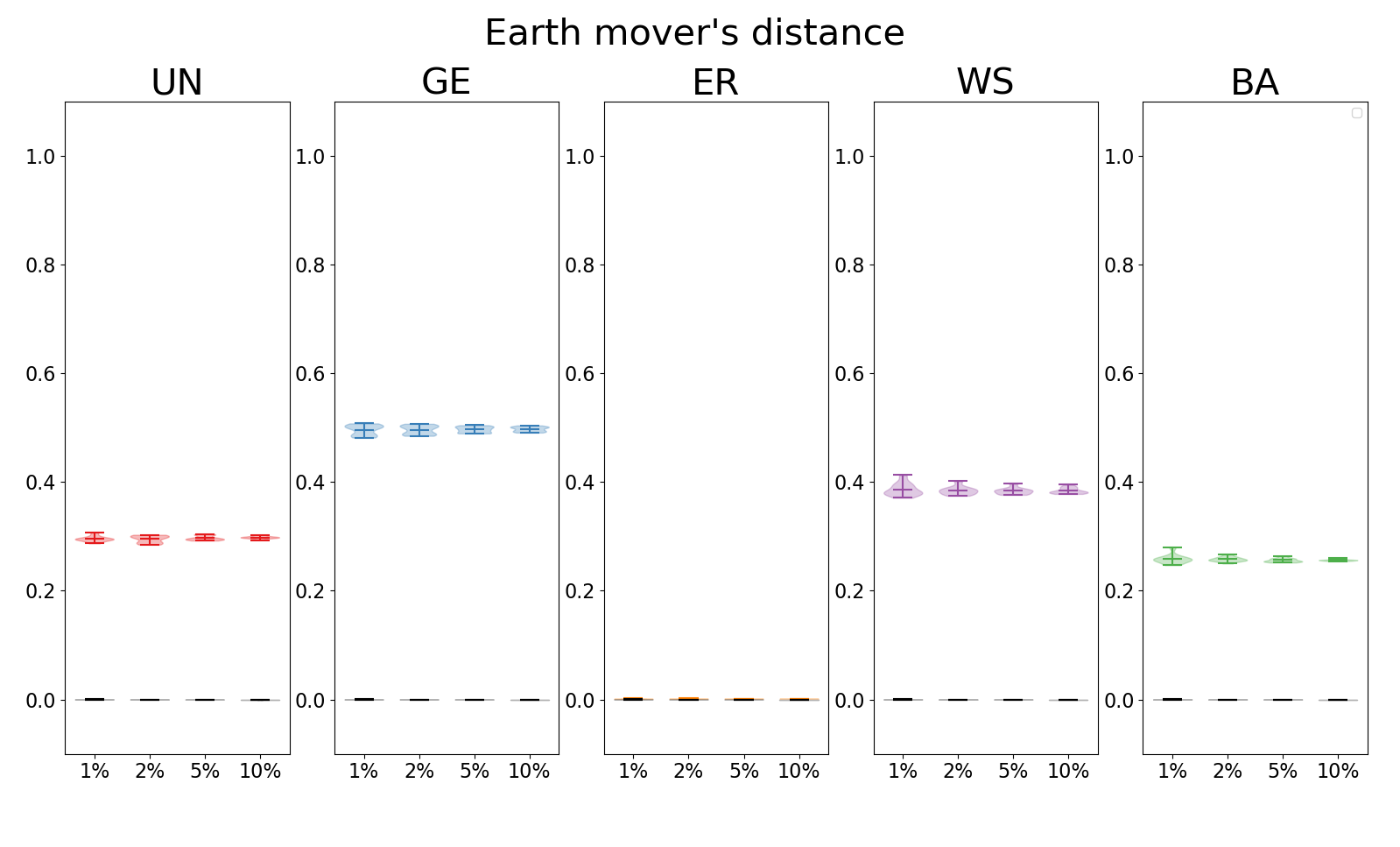}
 \caption{Comparison of stability with different sample sizes, 1\%, 2\%, 5\% and 10\% (with respect to the ground truth size) for graphs with $|V| = 9$ for two different representation measures and five different graph generators. The uniform sample from the ground truth data is shown in black.}% in all plots for comparison.}
 %each color represents, Noted, black violin represent {\color{green}benchmark}
 \label{fig:rep_benchmark}
 \end{figure*}
We compute the singular values of the dataset (similar to principal component analysis) and multiply them to obtain the \textit{robust ellipse} measure. As in the other approaches, we compute the robust ellipse measures for the generated sample and the ground truth dataset, calculate their volumes, and consider their ratio. 
This measure, unlike the diameter, bounding box and convex hull measures, should be more robust to outliers. One limitation of this measure is that it depends on the density of the dataset.
Note that for the representation measures, small values mean that the sample represents the ground truth well. For the coverage measures, the bigger the value, the better the coverage.
% That is, if the measure is close to $1$, it means that the sampled dataset covers the ground truth well.}

\section{Comparison Between Graph Generators}
\label{sec:rep_cov_exper}

In this section, we use the measures of representation and coverage defined in Sec.~\ref{sec:rep_cov} to compare between the five generators discussed in Sec.~\ref{sec:graph_generators}. In Sec.~\ref{sec:rep_results} we present results for the representation measure, in Sec.~\ref{sec:cov_results} we present results for the coverage measure, and in Sec.~\ref{sec:ba_ws_limitations} we discuss the limitations of graph generators.

\subsection{Representation of the Graph Generators} 
\label{sec:rep_results}

We start our analysis with Pearson correlations. For each of the five generators, we generate a sample with a size of $1\%$ of the ground truth dataset. We present the relative correlations for the sampled and ground truth datasets by calculating all pairwise 2D statistics of all non-isomorphic graphs with $|V| = 9$. While we computed these tables for all five generators, we only show the results for the best (ER) and worst (GE) performing generators, Fig. ~\ref{fig:correlations}. When comparing the ground truth and ER values, we see that they are nearly identical for all entries in the matrices and the largest difference is 0.09. When comparing the ground truth and GE, however, the differences are obvious and  as large as 1.00 (e.g., in the correlation between APL and density).
For the most representative model, ER, all pairwise correlations are similar for the ground truth and the sampled datasets. However, the graphs generated by ER do not cover the entire spectrum of possible values for each {\color{black}property}. %from the same figure we can also see that this generator does not cover the range of possible values of the ground truth (e.g., in the columns corresponding to APL, r, diameter and density, the leftmost and rightmost points in the plots are black).
%{\color{green}\color{green}Include the best and worst; show two figures; have two discussions.}

To effectively visualize the data, we use \textit{violin plots}~\cite{winter2012vioplot} (as implemented in the matplotlib library). Violin plots show more information than Box plots, as the kernel density visualization makes it possible to see more details about the data distribution (e.g., clusters). To make it easy to distinguish the results for the different generators, we use persistent colors across the visualizations and a color scheme from colorbrewer (Colorbrewer 2.0: \url{http://colorbrewer2.org/}).
%Sometimes the median and the mean are not enough to understand a dataset. With a box plot it is impossible to answer questions like: Are most of the values in the dataset clustered around the median or some percentile? Or are most of values clustered around the minimum and the maximum with nothing in the middle? Violin plot is the modification of the box plot which has two rotated kernel density on each side of the box. In order to have visually separable plots we use a color scheme from the colorbrewer (Colorbrewer 2.0: \url{http://colorbrewer2.org/}). 
\begin{table}[tbh]
\centering
\caption{The ranking (best to worst from left to right) of graph generators based on the representation measures KL and EM. It is notable that both the KL and EM measures give the same ranking of the 5 generators.  We are particularly interested in the best generators, and for $|V| = 7$, $|V| = 8$, $|V| = 9$ both KL and EM rank ER as best and BA as second best.}
\begin{tabular}{|l|l|l|l|l|l|}
\hline
$|V|$ & \multicolumn{5}{c|}{generators} \\ \hline
7 & ER   & BA   & UN   & WS   & GE  \\ \hline
8 & ER   & BA   & WS   & UN   & GE  \\ \hline
9 & ER   & BA   & UN   & WS   & GE  \\ \hline
\end{tabular}
\label{table:rep_perf_order}
\end{table}

% \begin{figure*}
% \centering
% \includegraphics[width=\columnwidth]{./representation_image/v=7_rep}
% \includegraphics[width=\columnwidth]{./representation_image/v=8_rep}

% \includegraphics[width=\columnwidth]{./representation_image/v=9_rep}
% \caption{Demonstration of the performance {\color{green}distance?} from 100\% sample size with ground truth for graphs with ($|V| = 7$ $|V| = 8$ $|V| = 9$) for two different representation measures and five different graph generators each color represents}
% \label{fig:rep_measure_vertex}
% \end{figure*}

Going beyond correlations, we also compute the KL divergence and the EM distance defined in Sec.~\ref{sec:representation} for samples generated by the five generators for $|V| = 7, 8, 9$. (We do not report results for $|V| = 5, 6$ as the sample sizes are too small and they do not contain enough information to calculate the KL divergence). % (often getting infinity values). 
In Fig.~\ref{fig:rep_measure_vertex}, we report the results of the $10$ different samples for each generator with sample size equal to the size of the ground truth dataset. As discussed in Sec.~\ref{sec:representation}, numbers close to 0 mean that the sample generated by the generator represents the ground truth well. The best generator for $|V| = 7, 8, 9$ is ER under both KL divergence and EM distance measures. Overall, GE, UN and WS are consistently outperformed by ER and BA.

\subsection{Consistency of the Representation Measure}
\label{sec:cons_repr}
An important question is whether the measures that we use (Sec.~\ref{sec:rep_cov}) behave consistently. Consistency  refers to being able to obtain \vahan{similar results for the representation measure} when the size of the ground truth data is changing, which happens (1) when we consider graphs with different number of vertices and (2) when we use different sample sizes.
We summarize the results from our consistency experiments, Fig.~\ref{fig:rep_measure_vertex}, in Table~\ref{table:rep_perf_order}. According to our experiments, KL and EM measures for representation are consistent for $|V| = 7$, $|V| = 8$ and $|V| = 9$.%, that is, they agree on the performance of all generators. 
%We would like to remark that in order to use the KL divergence we need a large enough sample, see the definition in Sec.~\ref{sec:representation}.
 
Another question is, how big does the sample need to be to represent the underlying ground truth dataset well. To answer this question, we run experiments for $|V| = 9$. For each of the five generators, we generate a sample with size $1\%$, $2\%$, $5\%$, and $10\%$ of the ground truth dataset. The results are reported in Fig.~\ref{fig:rep_benchmark}. We also show the results with a sample generated from the ground truth dataset by taking a uniform sample from it, which can be considered as a benchmark.
In Fig.~\ref{fig:rep_benchmark}, we see that ER performs as well as the uniform sample from the ground truth dataset. We also note that as the sample gets larger, the values for both KL divergence and EM distance get smaller (they are decreasing). This behavior is expected as the more graphs contained in the sample, the more representative they can be. Also worth noting is that the larger the sample size is, the less variation within the ten samples, as shown by the the progressively smaller violin plots.

\subsection{Coverage of the Graph Generators} 
\label{sec:cov_results}

\begin{figure*}[]
\includegraphics[width=2\columnwidth]{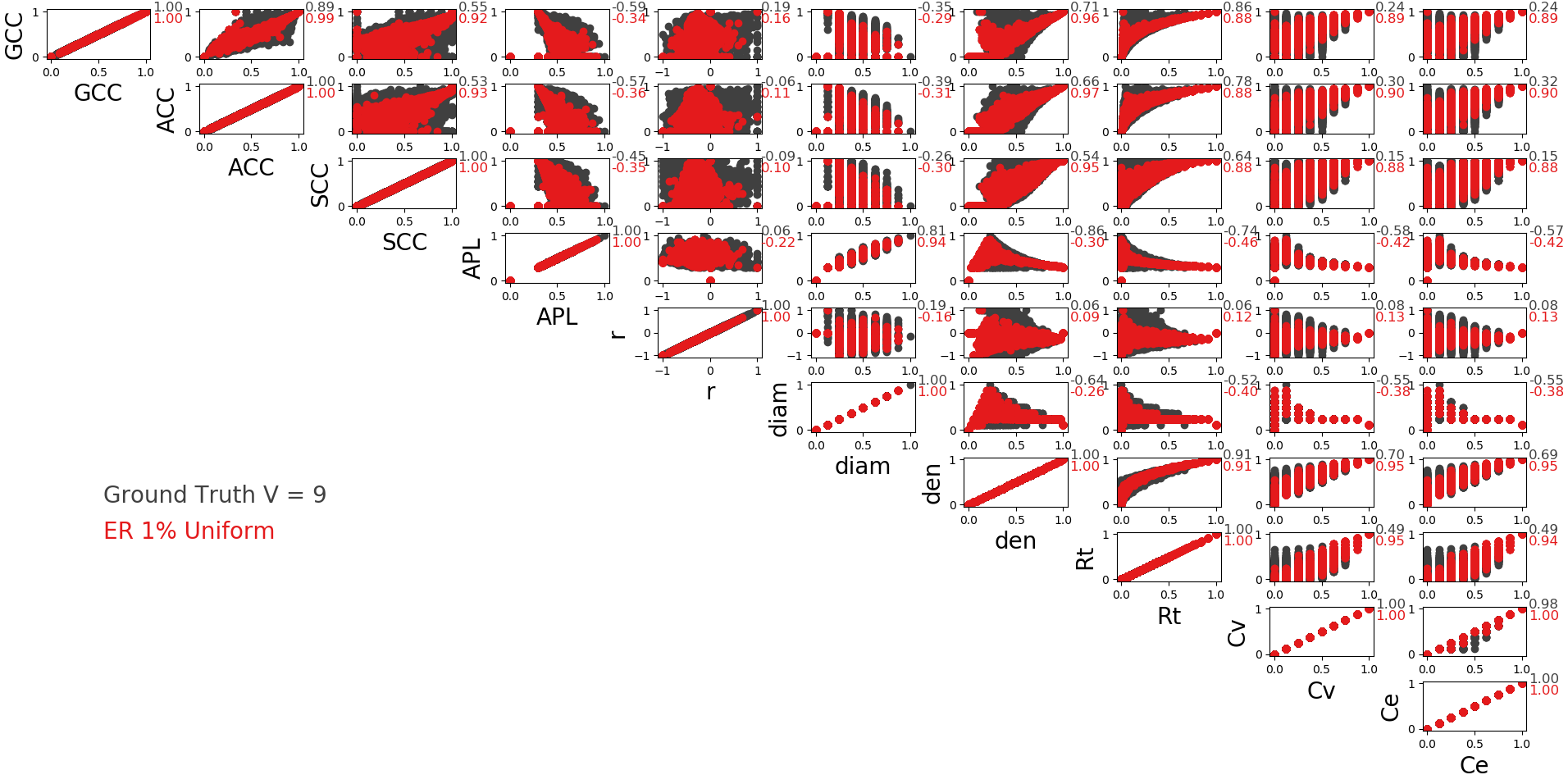}
\includegraphics[width=2\columnwidth]{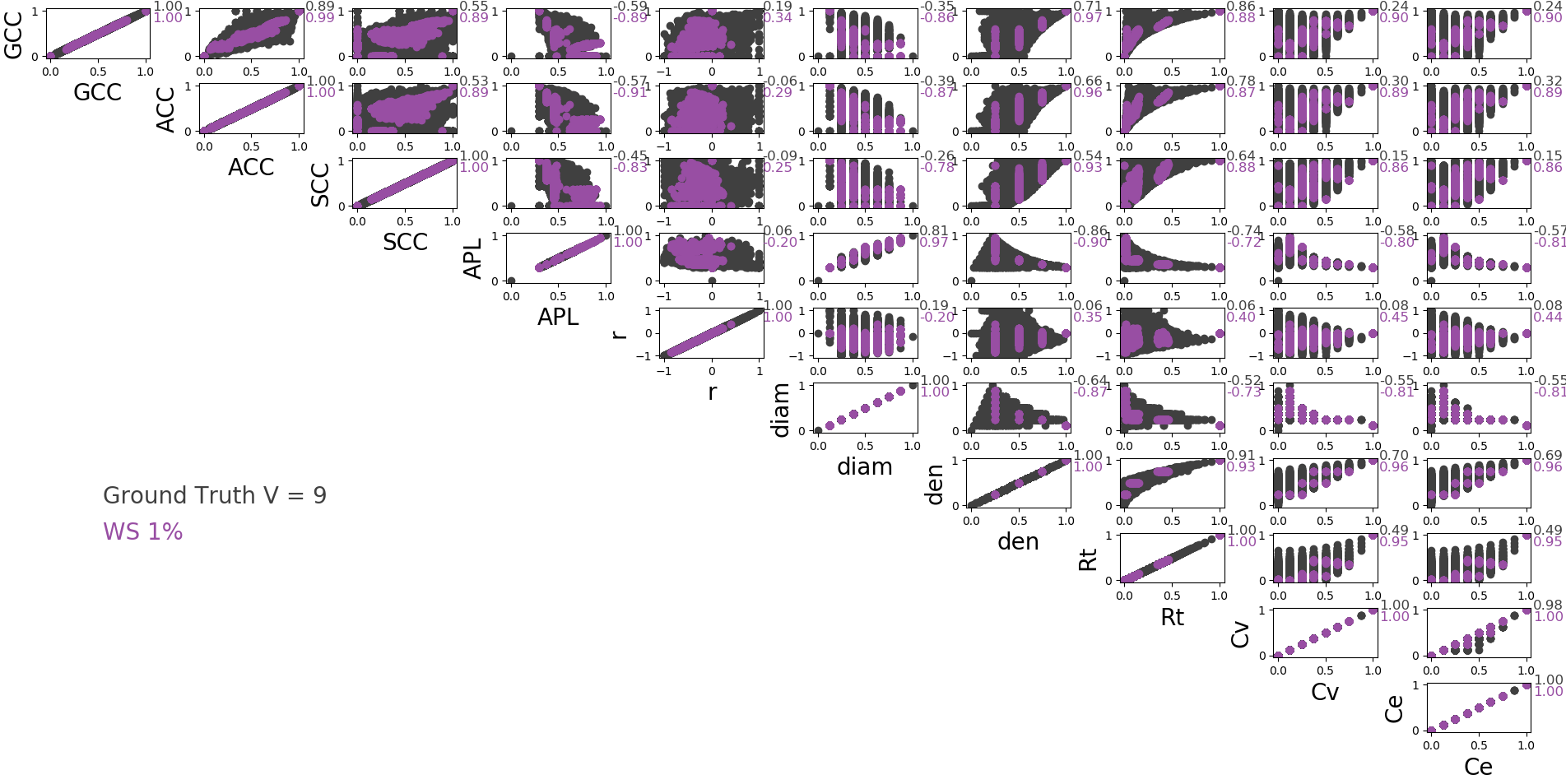}
\caption{Illustration of coverage for different generators based on 2D projections of the ground truth and generated data for $|V| = 9$. The ground truth is colored in black. The left figure shows a $1\%$ sample from the UN generator (red). The right figure shows a $1\%$  sample from the WS generator (purple). The corresponding correlation for the sample and the ground truth are shown next to each plot. Note that under each colored dot (red or purple) there exists a black dot, as the generators (red or purple) sample from the ground truth (black). Further, the fewer black points that are visible in the plot, the better the generator covers the ground truth.}
\label{fig:correlations_UNES}
\end{figure*}

We start our analysis of the coverage measures by visualizing the $2D$ plots between the ten graph properties; see Fig.~\ref{fig:correlations_UNES}. We only show the results for the UN and WS graph generators which achieve the best and worst coverage results, respectively. The ground truth is represented in black, the UN data in red (left) and the WS data in purple (right). 
%The left subfigure of Fig.~\ref{fig:correlations_UNES} presents the $2$D plots between the graph properties. In each plot the black dots correspond to the ground truth dataset and the red dots correspond to the dataset sampled (of size $1\%$ of the ground truth dataset) by the UN graph generator. Furthermore, the right subfigure of Fig.~\ref{fig:correlations_UNES} presents the $2$D plots between the graph properties. Similar to the left subfigure, the black dots correspond to the ground truth dataset and the purple dots correspond to the dataset (of size $1\%$ of the ground truth dataset) generated by the WS graph generator. 
Note that under each colored dot (red or purple) there exists a black dot, as the generators (red or purple) sample from the ground truth (black). Further, the fewer black points that are visible in the plot, the better the generator covers the ground truth. It is easy to see that UN performs much better, whereas WS misses large ranges of possible values, most noticeable in the density and assortativity columns (den and r).

\begin{figure*}[hp]
\centering
 \includegraphics[width = \columnwidth]{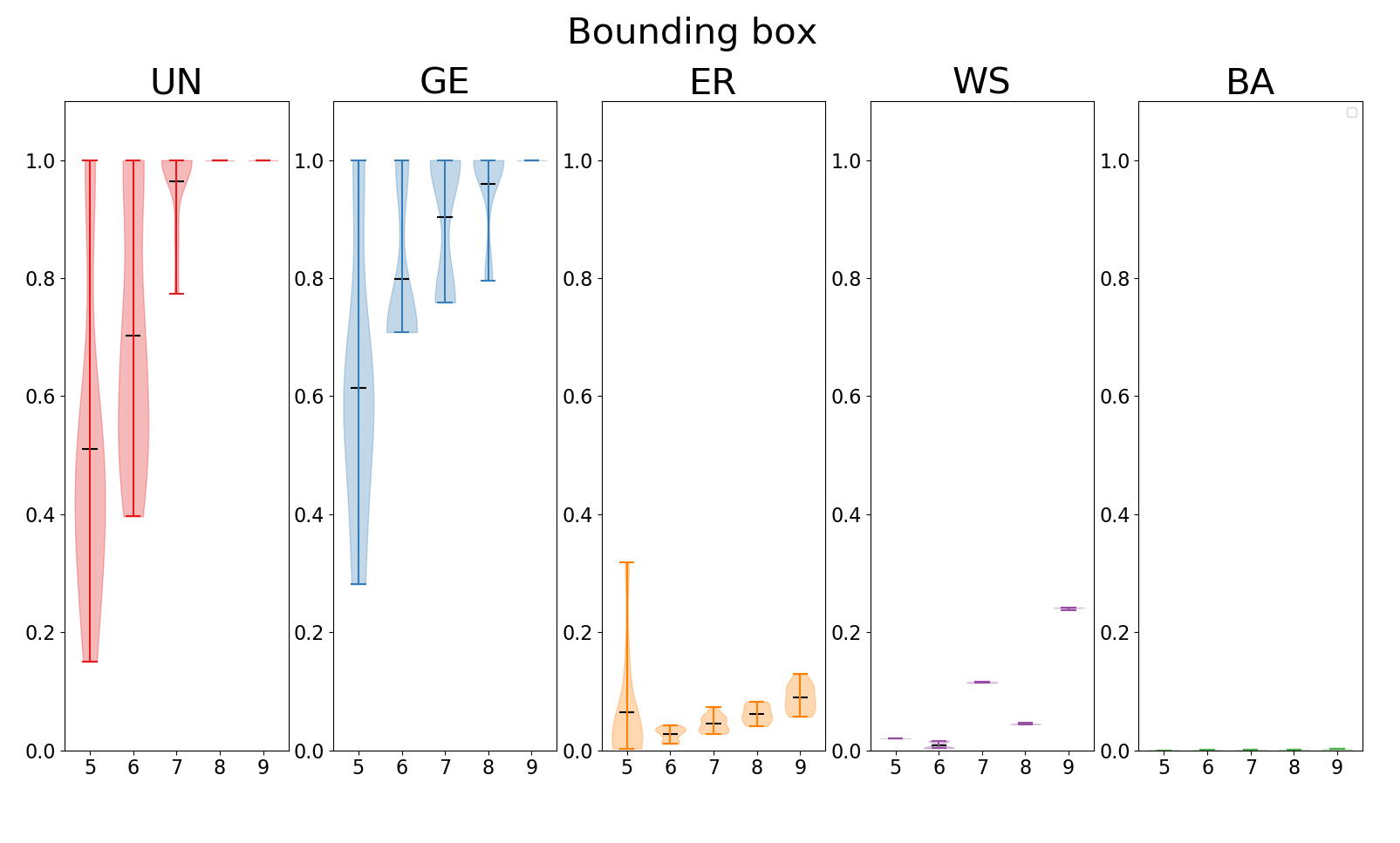}
 \includegraphics[width = \columnwidth]{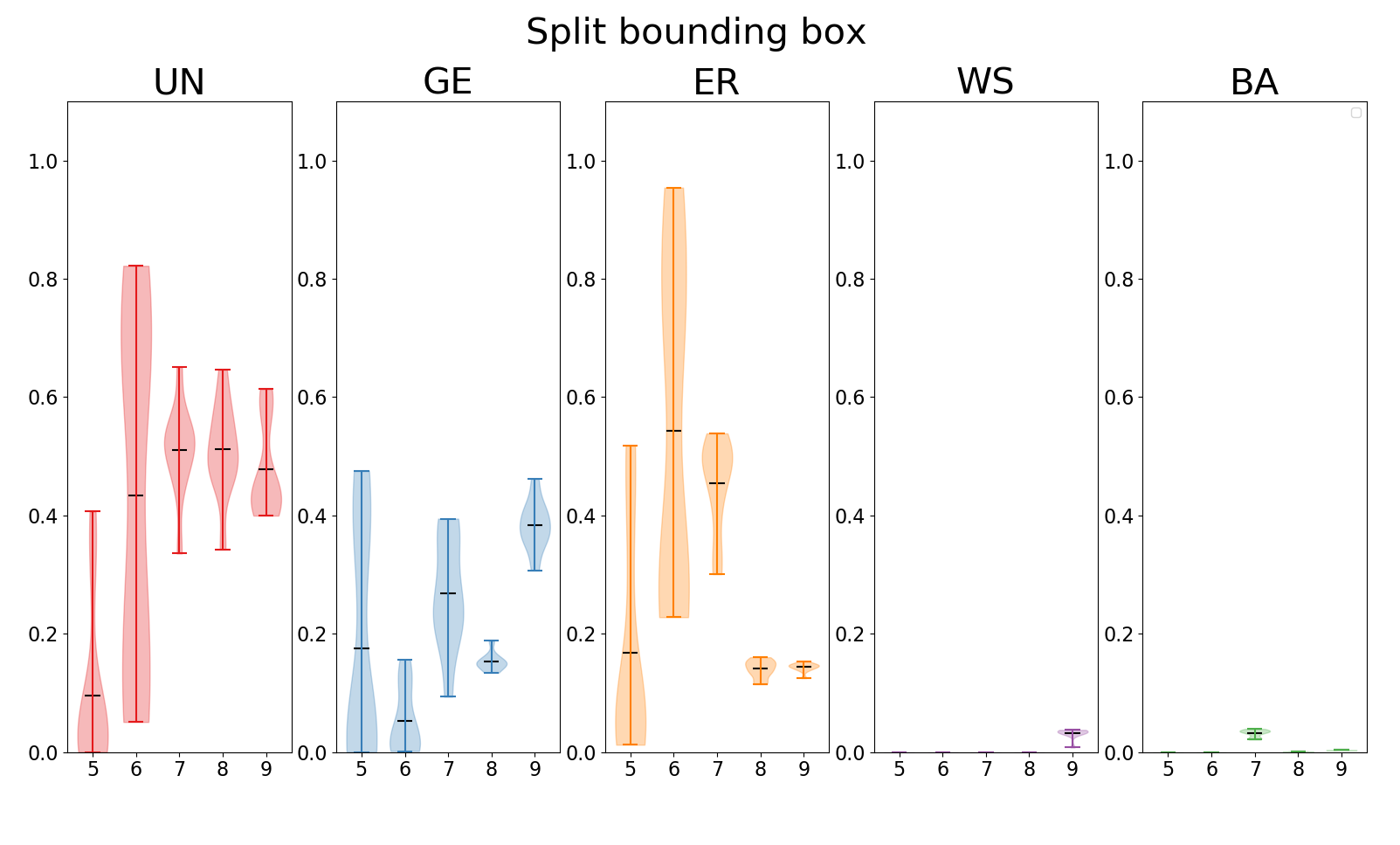}

 \includegraphics[width=\columnwidth]{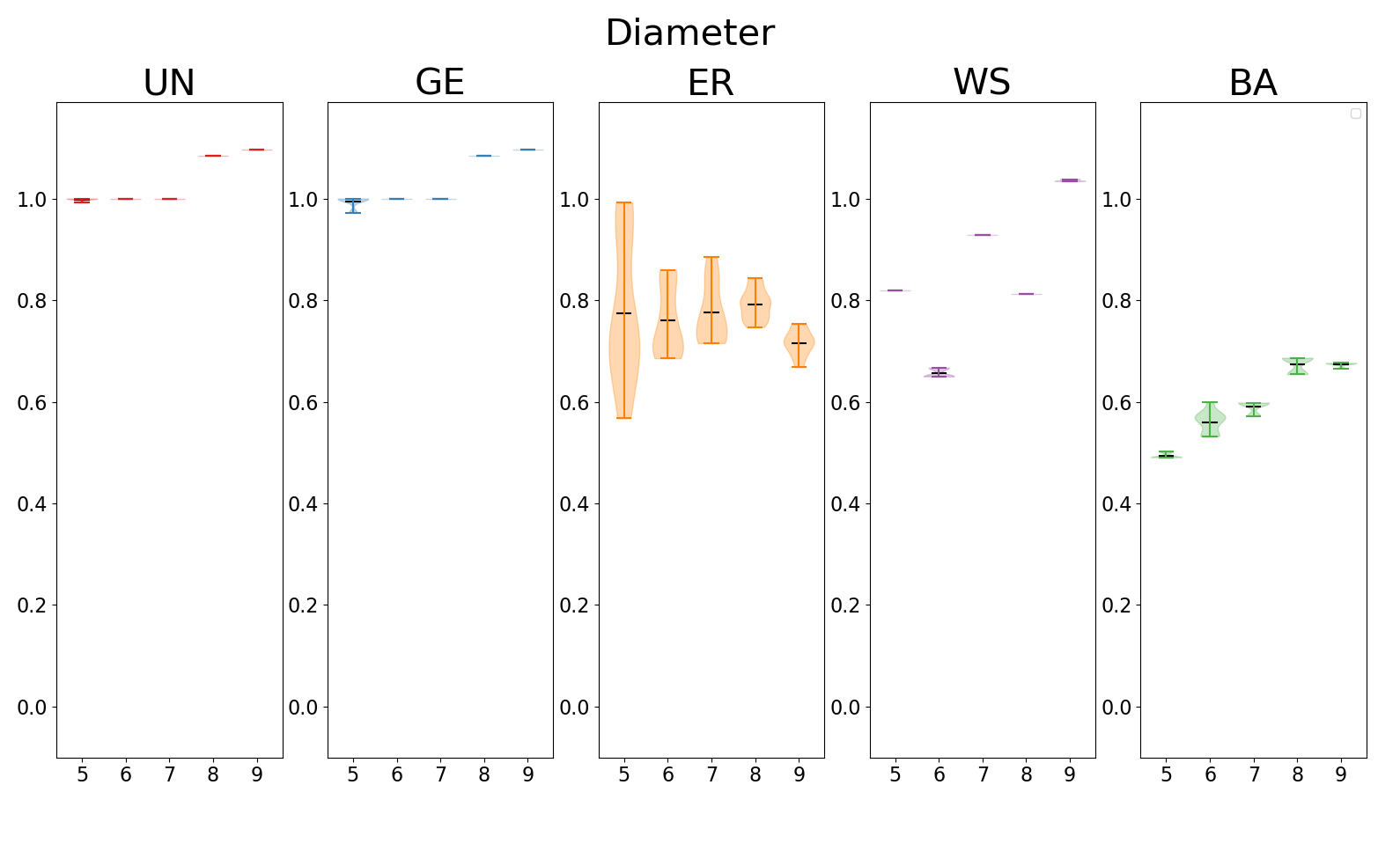}
 \includegraphics[width=\columnwidth]{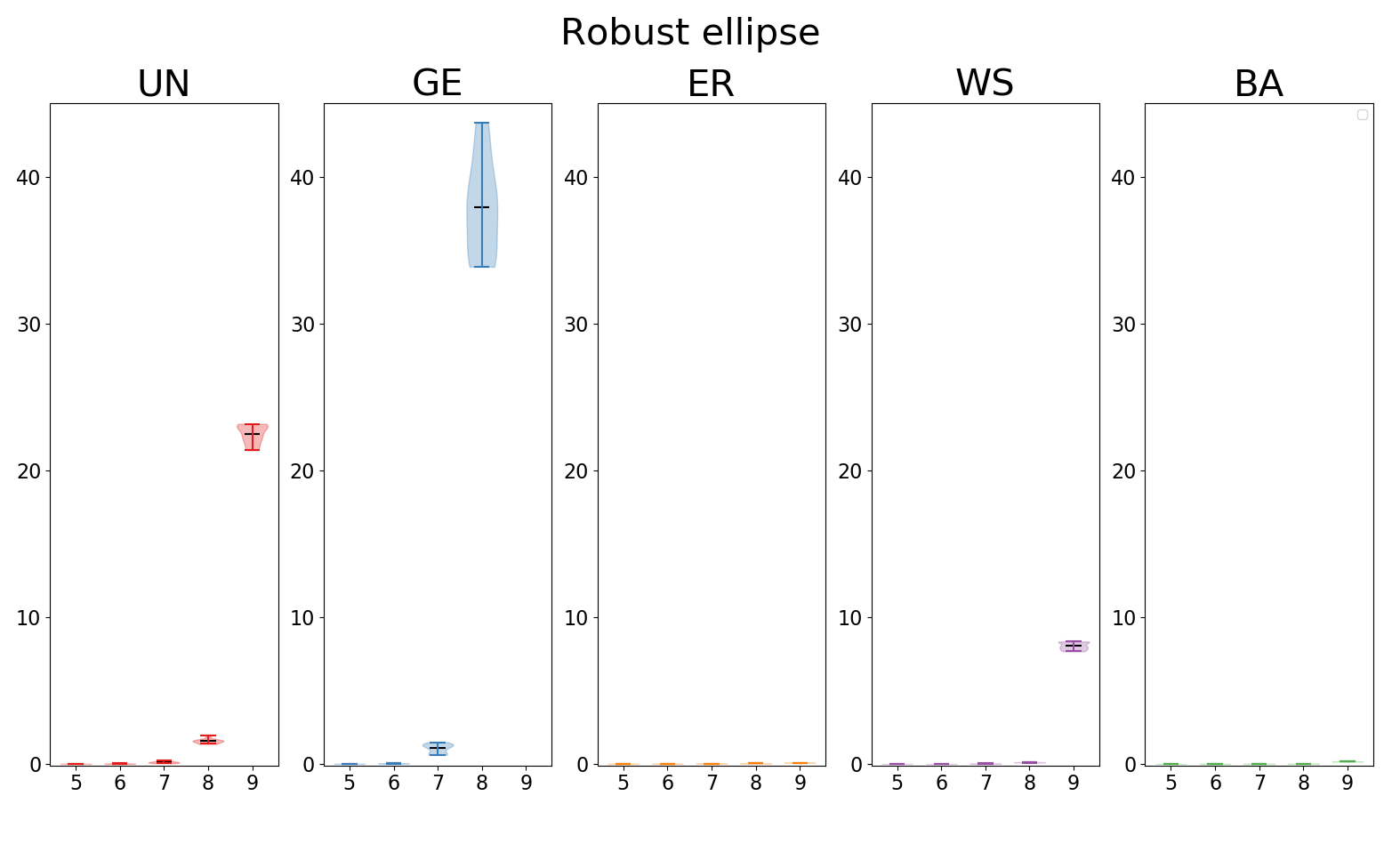}
\caption{Coverage comparison of the five graph generators (UN, GE, ER, WE and BA) over ten samples with sample sizes equal to the size of the ground truth dataset for $|V| = 5,6,7,8,9$. Each subfigure shows results for different coverage measures; see Sec.~\ref{sec:coverage}.}
% \caption{each plot represents a coverage measure, in each plot x axis represent $|V|$, y axis is the ratio of \Hang{measure(sample)/measure(ground truth)}}
 \label{fig:cov_measure_stable}
 \end{figure*}

Next, we use the four coverage measures discussed in Sec.~\ref{sec:coverage}. We run experiments for $|V|= 5, 6, 7, 8$, and $9$ with sample size equal to the ground truth dataset size, compute the four measures     for ten different samples and report the results using violin plots; see Fig.~\ref{fig:cov_measure_stable}. 
Unlike representation, where the ER-model was the best model, the ER-model performs poorly in coverage. 

Fig.~\ref{fig:cov_measure_stable} shows that for $|V| = 5, 6, 7, 8, 9$ the samples generated by the WS and BA models achieve poor coverage results, that is, the generated samples do not cover the range of graph  properties for the ground truth well. The samples generated by the UN and GE models achieve the best coverage (high values for bounding box, split-bounding box and diameter). This implies that the UN and GE cover the ground truth data best. Three of the measures, namely diameter, bounding box and robust ellipse give consistent results. The results for split bounding box do not always agree with the other three measures, which can be explained by taking into consideration two observations: (1) for ten dimensional data such as ours, the number of boxes jumps from 1 to 1024, and (2) for small values of $|V|$ there are not enough data points to calculate an accurate split bounding box measure.

%\Hang{ As what we discussed in section \ref{sec:coverage}, bounding box and diameter are hard to interpret when measures give a value close to 1. But robust ellipse does not have such problem. For example, GE have a better result compare to UN for $|V| = 9$ in robust ellipse while they are tied in bounding box and diameter. } \Hang{ For larger graphs set, robust ellipse may the most suitable one  over all measures we have for coverage.}

% \begin{figure*}
% \centering
% \includegraphics[width = \columnwidth]{./coverage_image/v=5}
% \includegraphics[width = \columnwidth]{./coverage_image/v=6}

% \includegraphics[width=\columnwidth]{./coverage_image/v=7}
% \includegraphics[width=\columnwidth]{./coverage_image/v=8}

% \includegraphics[width=\columnwidth]{./coverage_image/v=9}
% \caption{Comparison of the results of ... for graphs with ($|V| = 5$ $|V| = 6$ $|V| = 7$ $|V| = 8$) for four different coverage measures and four different graph generators each color represents}
% \label{fig:cov_measure_vertex}
% \end{figure*}

Note that UN and GE have worse results in terms of representation but the best results in terms of coverage. This is not particularly surprising as good representation and good coverage are correlated with different properties of the graph generators. For example, the UN and GE generators are more likely to create unusual/extreme graphs (fully connected, very sparse, etc.), whereas ER generates the most likely/typical graphs. 
We also note that if we use the strategy described in Sec.~\ref{sec:coverage} to approximate the diameter, in some cases the diameter of the ground truth dataset might be smaller than the diameter of the sampled dataset. We observe such behaviour in Fig.~\ref{fig:cov_measure_stable} and Fig.~\ref{fig:cov_benchmark}.

\subsection{Consistency of the Coverage Measure}
\label{sec:cons_cov}

As with the representation measure, we also consider the issue of consistency of the different coverage measures when we vary the sample size or the size of the ground truth data (by changing the number of vertices). Unlike the measures for representation (KL divergence and EM distance), some of the measures for coverage are not stable for small values of $|V|$. In Fig.~\ref{fig:cov_measure_stable}, we observe that for the bounding box and the split-bounding box measures, for $|V| = 5, 6, 7$ there is a high variation among the 10 samples. %This implies that for small graphs bounding box and split bounding box are not good measures to estimate the coverage.

\begin{figure*}[hp]
\centering
 \includegraphics[width = \columnwidth]{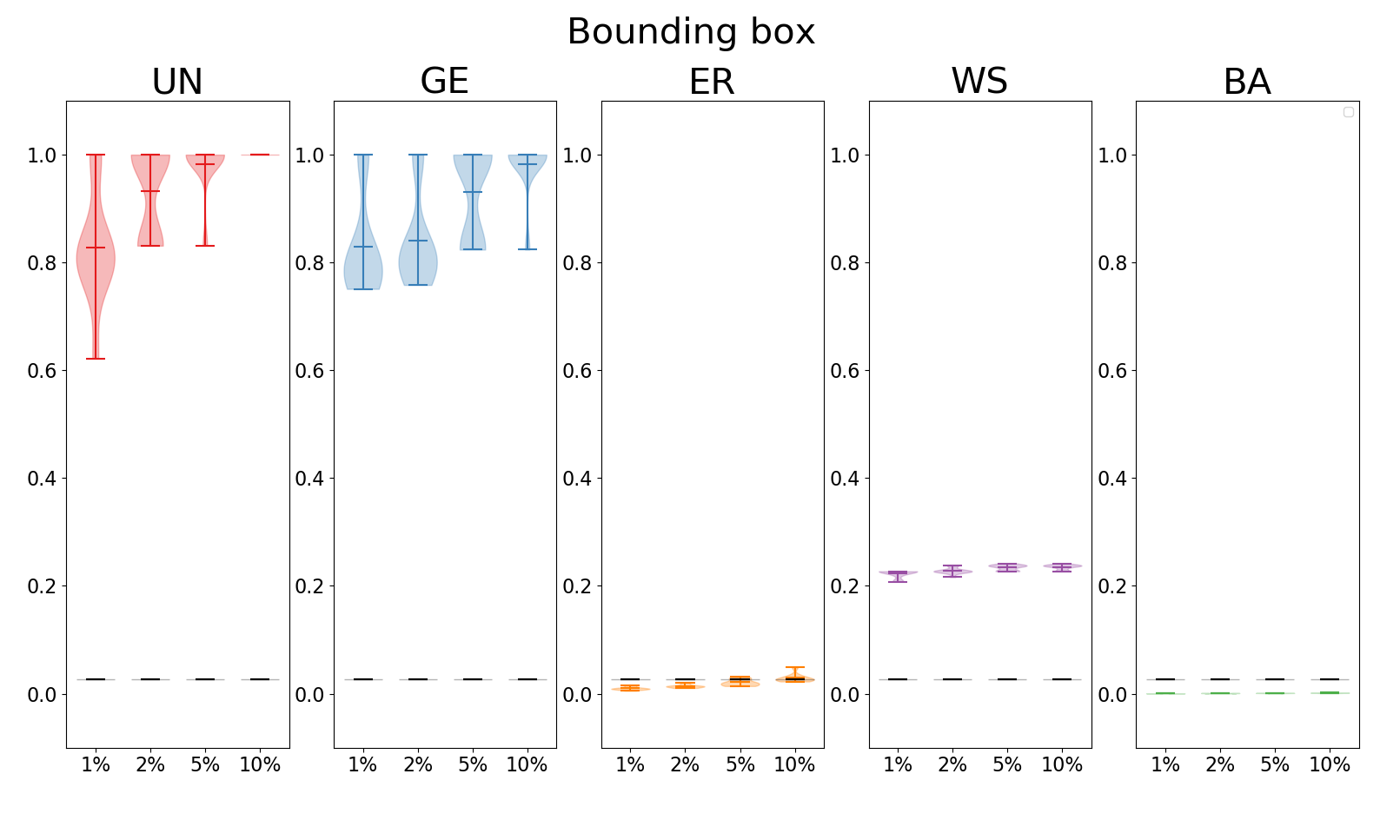}
 \includegraphics[width = \columnwidth]{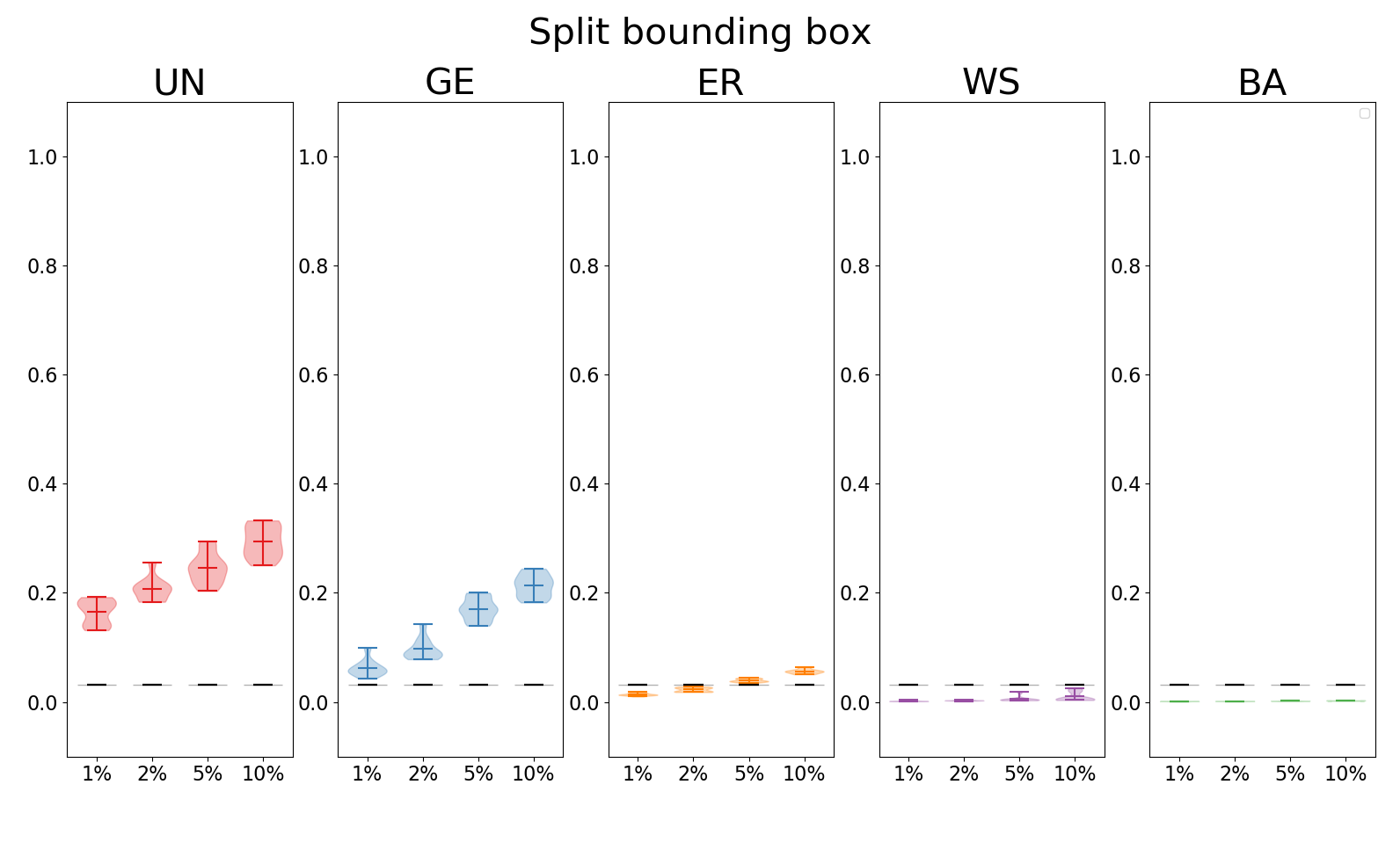}

 \includegraphics[width=\columnwidth]{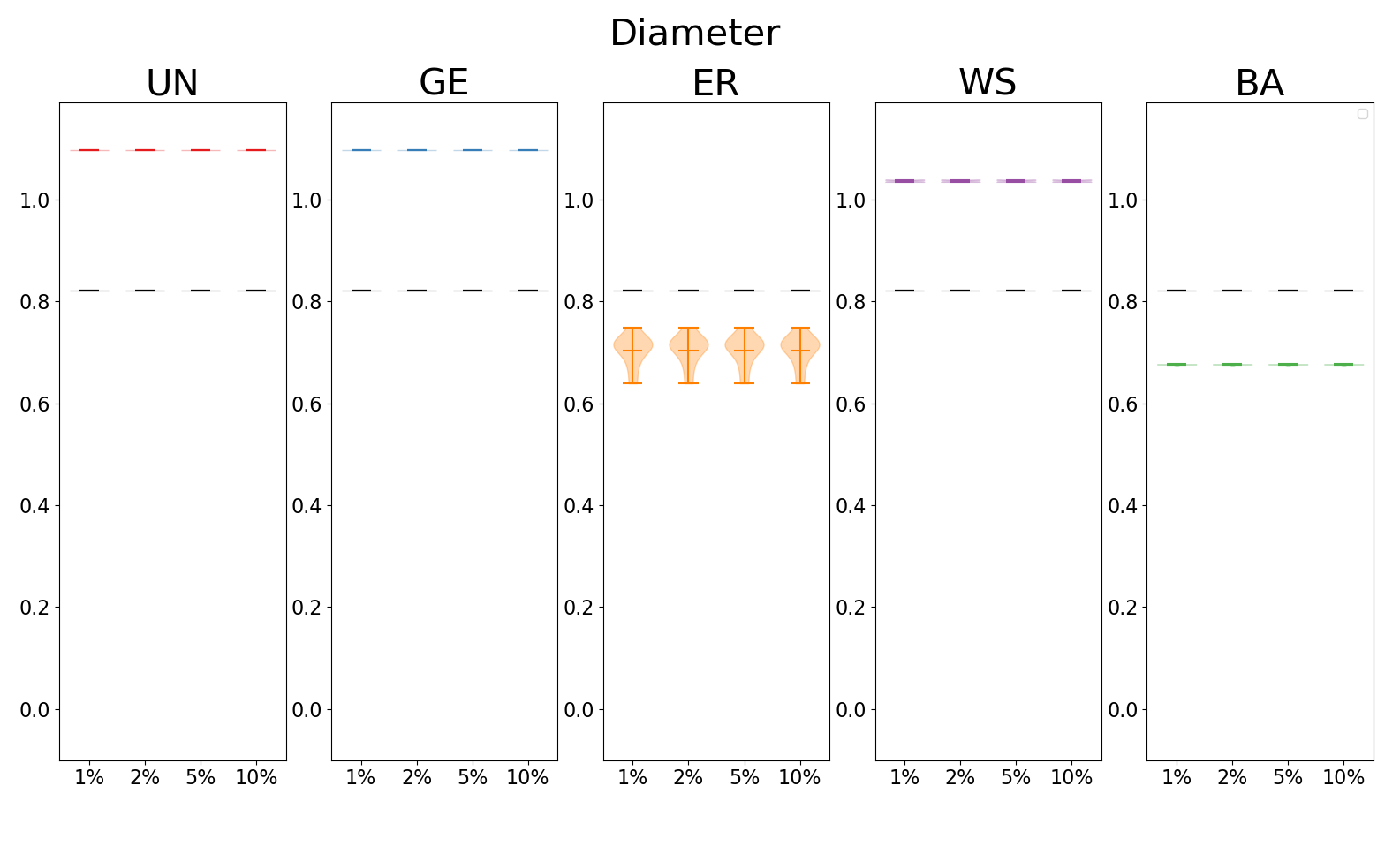}
 \includegraphics[width=\columnwidth,height=5.4cm]{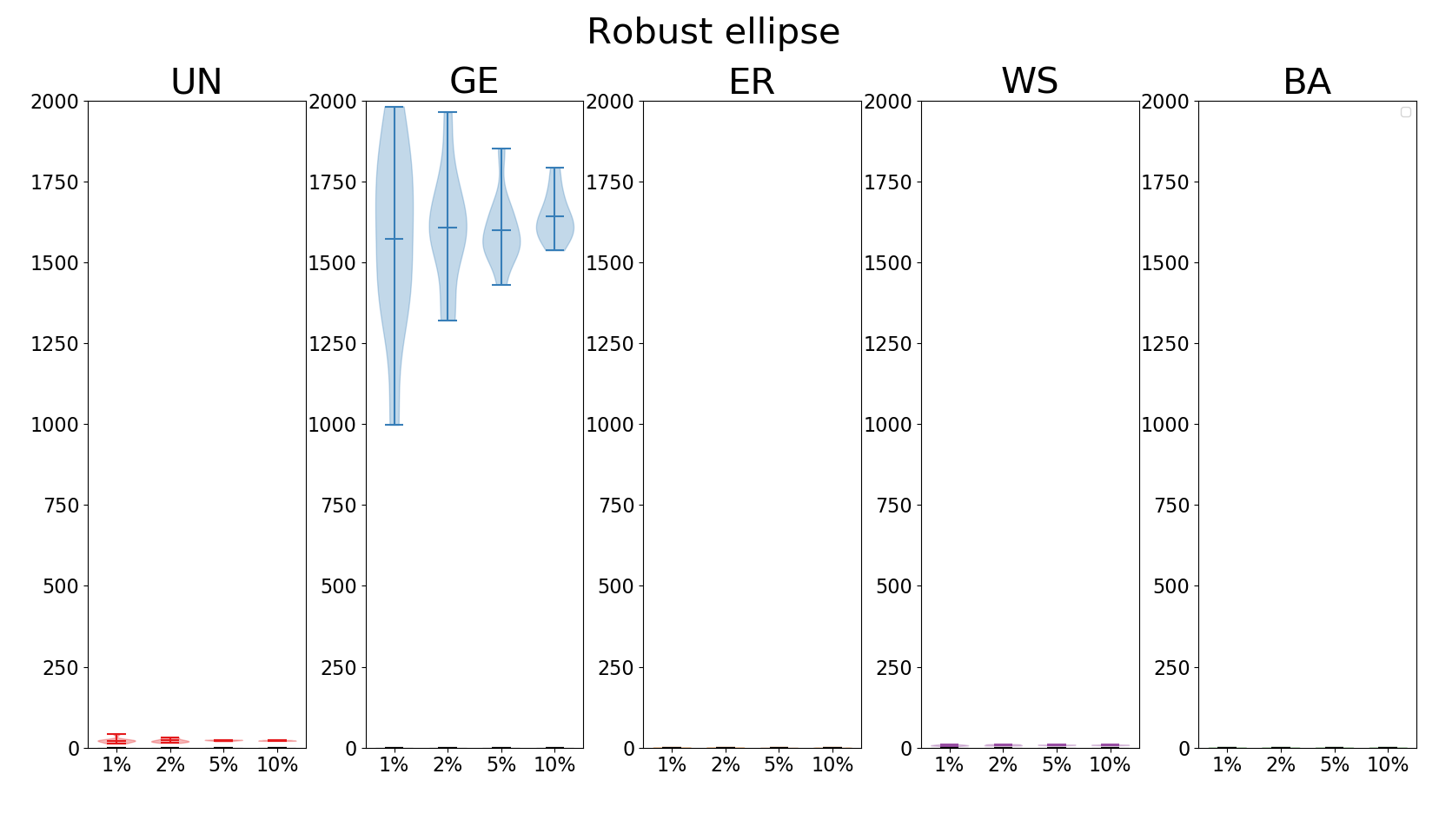}
\caption{Comparison of stability when using different sample sizes, 1\%, 2\%, 5\% and 10\% (with respect to the ground truth size) for graphs with $|V| = 9$ for four different coverage measures and five different graph generators. The uniform sample from the ground truth data is shown in black in all plots.}
% \caption{each plot represents a coverage measure, in each plot x axis represent $|V|$, y axis is the ratio of \Hang{measure(sample)/measure(ground truth)}}
 \label{fig:cov_benchmark}
 \end{figure*}

We also observe some variation in the diameter, but only for the ER generator; see Fig.~\ref{fig:cov_measure_stable} and Fig.~\ref{fig:cov_benchmark}. However, this is an expected behaviour, since the UN and GE graph generators are able to capture the extreme cases. Thus, even for small samples, the ratio of diameters of the sample and the ground truth is close to $1$. However, for the ER graph generator, for smaller samples the generator does not always capture the extreme cases.
%, thus there is variation.
 
\subsection{Limitations of BA and WS}
\label{sec:ba_ws_limitations}

None of the graph generators explicitly optimize representation or coverage, but some are  better than others. As shown in Figures~\ref{fig:rep_measure_vertex}, \ref{fig:rep_benchmark}, \ref{fig:cov_measure_stable} the BA and WS generators perform poorly across all representation and coverage measures. The underlying generation methods and the specific parameter settings used might explain why. The WS small-world graph generator requires 3 parameters: the number of vertices $n$, a number $k$ that specifies how many neighbors each node should be connected to, and a probability $p$ for adding these edges. For our experiments we used $n = 9$, $k$ chosen uniformly at random in the range $2$ to $|V|-1$, and $p$ chosen uniformly at random in the range $[0, 1]$. Since the WS generator begins with a $k$-connected ring, and only switches an edge from one node to another, the result is graphs with $k*n$ edges, which limits the number of non-isomorphic graphs. 

Similarly, the BA generator requires 2 parameters: the number of vertices $n$ and the number of edges $m$ to attach from a new node to existing nodes. We use $n = 9$ and randomly chose an integer value from $1$ to $|V|-1$ for $m$. Thus, the possible number of edges can only be $(n-m)*m$, which in the case of $|V| = 9$ leads to only four possible values of $|E|$  $(8, 14, 18, 20)$. This restricts the range of different graphs that can be generated.
 
%\begin{figure}[t]
%\includegraphics[width=\linewidth]{./images/ER_Population}
%\caption{Ground truth (blue) and population edge distribution for ER (red).}
%\label{fig:P7to10}
%\end{figure}

%From the correlation matrix for ER using the population distribution we can see that correlations are close to the ground truth; see Fig.~\ref{fig:P7to10}. The correlations are worse when we use uniform distribution, but the coverage of the extreme cases is much better; see Fig.~\ref{fig:U7to10}.  This is especially evident from the columns corresponding to APL, diameter and Rt where the leftmost and/or rightmost points in the plots are blue in the population experiment and red in the uniform.

%\begin{figure}[h]
%\includegraphics[width=\linewidth]{./images/ER_Uniform}
%\caption{Ground truth (blue) and uniform at random edge distribution for ER (red).}
%\label{fig:U7to10}
%\end{figure}

\section{Conclusions and Future Work}
\label{sec:conclusion}

We discussed an exploration of the space of graphs, treating each graph as a high dimensional object, where the number of dimensions is determined by the number of global properties associated with each graph. For low-order graphs ($|V|\leq 10$) and for a small number of fixed properties (10 in this paper) we can directly explore the space. This allows us to find many instances of the ``same stats, different graphs" phenomenon. One natural question is to determine the ``true" set of dimensions for the space of non-isomorphic graphs, where we consider a set of graph properties that suffice to always distinguish two non-isomorphic graphs. 

For larger graphs we can use random graph generators. Since most practical random graph generators sample the space of isomorphic graphs, they are not equally well-suited for the task of generating non-isomorphic graphs. We considered two particular measures, coverage and representation, and find that UN does a good job in coverage and ER is best for representation. 
%For the purpose of studying graph properties and structure, we need generators that represent and cover the space of non-isomorphic graphs. 
%According to our experiments, there are no $2$ non-isomorphic graphs with $|V| < 7$ that share the same exact $10$ properties, defined in Sec.~\ref{sec:graph_stats}. However, as $|V|$ gets larger we observe that there are some non-isomorphic graphs that share the same exact $10$ properties, for $|V| = 7$ there is just one pair, for $|V| = 8$ there are 8 pairs, for $|V| = 9$ there are even triples,
%\hang{ for $|V| = 10$ there are 5 quintuple(5-tuples)} \hang{I also update repetition\_graph\_set\_num.txt in the folder, it contians detail. in case we plan to describe more}. 
%Even if we increase the number of properties recorded for each graph we expect that for large number of vertices (large values of $|V|$) there will be many ``same stats, different graphs." 
%computed, for larger that can capture the differences of graphs that the current $10$ properties are not able to capture.
We provide online exploration tool, source-code and data at \url{http://differentgraphs.cs.arizona.edu}

%We considered how to explore the space of graphs and graph properties that make it possible to have multiple graphs that are identical in a number of graph properties, yet are clearly different. 

To ``see" the difference between two non-isomorphic graphs, it often suffices to look at the drawings of the graphs. However, as graphs get larger, some graph drawing algorithms may not allow us to distinguish differences in properties between two graphs purely from their drawings. 
We recently studied how the perception properties, such as density and ACC, is affected by different graph drawing algorithms~\cite{soni2018perception}. The results confirm the intuition that some drawing algorithms are more appropriate than others in aiding viewers to perceive differences between underlying graph properties. Further work in this direction might help ensure that differences between graphs are captured in the different drawings.
Other interesting open problems that arise from this work include: 
%Finding more detailed analysis on finding different graphs with the same stats. Finding 
designing efficient generators for graphs with the same stats or graphs that share some stats and vary in others, finding theoretical guarantees and bounds on how likely it is to find such graphs for specific generators, computing bounds on how many graphs have the same subset of stats.

% Can use something like this to put references on a page
% by themselves when using endfloat and the captionsoff option.
\ifCLASSOPTIONcaptionsoff
  \newpage
\fi

% trigger a \newpage just before the given reference
% number - used to balance the columns on the last page
% adjust value as needed - may need to be readjusted if
% the document is modified later
%\IEEEtriggeratref{8}
% The "triggered" command can be changed if desired:
%\IEEEtriggercmd{\enlargethispage{-5in}}

% references section

% can use a bibliography generated by BibTeX as a .bbl file
% BibTeX documentation can be easily obtained at:
% http://mirror.ctan.org/biblio/bibtex/contrib/doc/
% The IEEEtran BibTeX style support page is at:
% http://www.michaelshell.org/tex/ieeetran/bibtex/
\bibliographystyle{IEEEtran}
% argument is your BibTeX string definitions and bibliography database(s)
%\bibliography{IEEEabrv,reference}
% Generated by IEEEtran.bst, version: 1.14 (2015/08/26)

%
% <OR> manually copy in the resultant .bbl file
% set second argument of \begin to the number of references
% (used to reserve space for the reference number labels box)

%%use following if all content of bibtex file should be shown
%\nocite{*}

\begin{IEEEbiographynophoto}{Hang Chen} is a graduate student at the Department of Computer Science in the University of Arizona. He received a BS degree in Mathematics and Computer Science from the University of Arizona. His research interests include graph theory, information visualisation, and machine learning.
\end{IEEEbiographynophoto}

\begin{IEEEbiographynophoto}{Utkarsh Soni} is a PhD student in the School of Computing, Informatics, and Decision Systems Engineering at Arizona State University. He received a Masters in Computer Science at Arizona State University. His Master's thesis was on Graph Perception. His current research interests are human-aware AI and human preference modeling.
\end{IEEEbiographynophoto}
% \vfill

\begin{IEEEbiographynophoto}{Yafeng Lu} is a software engineer at Bloomberg L.P. Before joining Bloomberg, she was a postdoctoral research associate in the School of Computing, Informatics, and Decision Systems Engineering at Arizona State University.  She received her PhD degree from ASU.  Her research interests include predictive visual analytics, social media and textual data analysis.
\end{IEEEbiographynophoto}

\begin{IEEEbiographynophoto}{Vahan Huroyan} is a postdoctoral research associate at the Department of Mathematics at the University of Arizona and is affiliated with the NSF Tripods research group. He received his PhD degree in mathematics from University of Minnesota. His research interests include machine learning, mathematical data analysis, computer vision, and non-convex optimization.
\end{IEEEbiographynophoto}

\begin{IEEEbiographynophoto}{Ross Maciejewski} is an Associate Professor in the School of Computing, Informatics, and Decision Systems Engineering at Arizona State University. He received his PhD degree from Purdue University. His research interests include visual analytics, geographic visualization, and data science.
\end{IEEEbiographynophoto}

\begin{IEEEbiographynophoto}{Stephen Kobourov} is a Professor at the Department of Computer Science at the University of Arizona. He received a BS degree in Mathematics and Computer Science from Dartmouth College and MS and PhD degrees from Johns Hopkins University. His research interests include information visualisation, graph theory, and geometric algorithms.
\end{IEEEbiographynophoto}

\vfill
% \enlargethispage{-5in}
\end{document}